\documentclass[11pt]{article}
\usepackage{arxiv}
\renewcommand{\shorttitle}{A Common Lower Bound for Dantzig and Modified Simplex}

\usepackage[numbers,square,sort&compress]{natbib}
\usepackage[T1]{fontenc}
\usepackage[utf8]{inputenc}
\usepackage{amsmath,amssymb,amsthm,mathtools}
\usepackage{booktabs,array,tabularx,enumitem,microtype,placeins}
\usepackage{algorithm,algpseudocode}
\usepackage{xcolor}
\usepackage{tikz}
\usetikzlibrary{arrows.meta}
\usepackage[
  colorlinks=true,
  citecolor=blue,
  linkcolor=blue,
  urlcolor=blue
]{hyperref}

\hypersetup{
  pdftitle={A Common Exponential Lower Bound for Dantzig and Modified Simplex in Deterministic Discounted Games},
  pdfauthor={Xinyu Qiao, Tao Yao}
}

\allowdisplaybreaks[2]
\theoremstyle{plain}
\newtheorem{theorem}{Theorem}[section]
\newtheorem{lemma}[theorem]{Lemma}
\newtheorem{proposition}[theorem]{Proposition}
\newtheorem{corollary}[theorem]{Corollary}
\theoremstyle{definition}
\newtheorem{definition}[theorem]{Definition}
\theoremstyle{remark}
\newtheorem{remark}[theorem]{Remark}
\newcommand{\R}{\mathbb R}
\newcommand{\Z}{\mathbb Z}
\newcommand{\A}{\mathcal A}

\newcommand{\rank}{\operatorname{rank}}
\newcommand{\src}{\operatorname{src}}
\newcommand{\Max}{\mathrm{Max}}
\newcommand{\Min}{\mathrm{Min}}
\newcommand{\orig}{\mathrm{orig}}
\newcommand{\ret}{\mathrm{ret}}
\newcommand{\trav}{\mathrm{travel}}
\newcommand{\ent}{\mathrm{enter}}
\newcommand{\skp}{\mathrm{skip}}
\newcommand{\brd}{\mathrm{board}}
\newcommand{\sty}{\mathrm{stay}}
\newcommand{\lev}{\mathrm{leave}}
\newcommand{\norm}[1]{\lVert#1\rVert_\infty}
\tikzset{edge/.style={-{Stealth[length=2mm]},semithick},
  state/.style={circle,draw,minimum size=9mm,inner sep=2pt},
  gate/.style={rectangle,draw,minimum size=9mm,inner sep=3pt}}
\title{A Common Exponential Lower Bound for Dantzig and Modified Simplex in Deterministic Discounted Games}
\author{Xinyu Qiao\hspace{1.5em}Tao Yao}
\date{}
\begin{document}
\maketitle

\begin{abstract}
On a family of $N$-state deterministic two-player discounted games, every run that chooses adaptively between Dantzig's rule and modified simplex requires $2^{\Omega(N)}$ switches from a specified initial strategy, regardless of tie breaking. More generally, a return-gate construction lets Dantzig reproduce any finite Bland prefix in an undiscounted deterministic Markov decision process with rational rewards whenever every visited policy reaches a common sink from all states. Applied to an acyclic binary counter, deleting the minimizer's return actions preserves every fixed maximizing strategy's value at each original state while reducing Dantzig's switch count from exponential to linear. With returns, halving the total optimality gap can take $\Omega(h/\log h)$ switches at discount horizon $h=(1-\gamma)^{-1}$, even on games solvable in linear arithmetic time. Conversely, Dantzig is strongly polynomial when strongly connected components have bounded size, each state has a bounded number of actions, and the condensation graph has height $O(\log N)$.
\end{abstract}

\section{Introduction}

The running time of strategy iteration depends on both the game model and the rule used to select improving actions. For deterministic discounted Markov decision processes (MDPs), Dantzig's rule switches the policy at one state to an action with the largest positive advantage and then reevaluates the policy. \citet{post2015simplex} prove that this algorithm runs in strongly polynomial time independently of the discount factor, and \citet{hansen2014dantzig} sharpen their pivot bound by a factor of the number of states. Does this guarantee extend to deterministic two-player games when each strategy evaluation incorporates an exact best response from the opponent?

Fixing the strategy of the maximizing player (Max) leaves the minimizing player (Min) with a deterministic MDP, so computing one best response takes strongly polynomial time. Even so, Max can be forced through exponentially many switches. In the single-player analysis of \citet{post2015simplex}, repeated traversals of a newly created cycle amplify the gain from switching into it. In a two-player game, however, even when the old minimizing response keeps play on a profitable cycle after Max switches, Min's updated best response can exit that cycle and erase the amplified gain. In concurrent and independent work, \citet{mei2026simplestrategyiterationdantzigsrule} exploits this mechanism to build a binary counter on which Dantzig makes exponentially many switches.

A return-gate construction isolates the role of Min's response. Inserting a minimizing gate on each edge, where Min can either proceed forward or send play back to the starting state, preserves the value of every fixed Max strategy at every original state, yet reshapes the current advantages that guide Dantzig's next switch. From the same initial strategy on $k$ levels, Dantzig makes only $k$ switches when Min must go forward, versus $2^{k+1}-k-2$ when Min can return (Proposition~\ref{prop:forward-only}). Can Max avoid such long trajectories by scoring each candidate switch by its actual gain after Min reoptimizes? A second family rules this out: it forces exponentially many switches for the modified simplex method of \citet{jia2020towards}, and indeed for any rule that chooses adaptively between modified simplex and Dantzig.

\subsection{Main contributions}

\textbf{A common exponential lower bound.} At each iteration, Max may apply either Dantzig's rule, maximizing the current one-step advantage, or modified simplex, maximizing the total value gain across all states after Min reoptimizes. Theorem~\ref{thm:common-lower-bound} constructs a single family of deterministic discounted games on which every such run makes the same exponential number of switches, even when the choice of rule depends on the entire history and ties are broken arbitrarily. In this construction, each switch of an underlying binary counter forces two consecutive switches in the game, one at each of two copies of the affected state. Extending this replication to $d+1$ copies per state yields an exponential switch lower bound for fixed-depth lookahead of any depth $d>1$ under a last-label tie-breaking rule (Theorem~\ref{thm:lookahead-exponential}).

\textbf{From Bland to Dantzig.} Given any finite Bland trajectory prefix in an undiscounted deterministic MDP with rational rewards in which every visited policy reaches a common sink from all states, Theorem~\ref{thm:transfer} constructs a deterministic two-player discounted game in which Dantzig reproduces the prefix step for step, uniquely selecting the lift of Bland's lowest-ranked improving action at each iteration. Applying this transfer to an acyclic counter adapted from \citet{disser2025lower} yields the Dantzig lower bound in Theorem~\ref{thm:small-scc}, and Section~\ref{sec:source} further illustrates the transfer on a cyclic counter.

\textbf{Time to halve the optimality gap.} Let $h=(1-\gamma)^{-1}$ denote the effective horizon at discount $\gamma$. For every sufficiently large rational $h$, the family in Theorem~\ref{thm:small-scc} contains games with $\Theta(\log h)$ states and actions on which Dantzig requires $\Omega(h/\log h)$ switches just to halve the sum of statewise optimality gaps (Corollary~\ref{cor:small-scc-progress}), nearly matching the general $O(h\log h)$ halving upper bound at this scale.

\textbf{Strongly polynomial solution beyond forward games.} In the forward games of \citet{mei2026simple}, every strongly connected component (SCC) is controlled by a single player. Theorem~\ref{thm:component-solver} extends their backward component-by-component solver to mixed SCCs (components containing states of both players), provided each Min state in a mixed SCC has at most one action remaining inside that component. This algorithm computes optimal values at all states and an optimal stationary strategy pair in $O(N^2M)$ arithmetic operations, where $M$ is the total number of actions. All lower-bound games above belong to this class, and the Dantzig instances in Theorem~\ref{thm:small-scc} are solvable in $O(N+M)$ arithmetic operations (Corollary~\ref{cor:small-scc-linear}).

\textbf{Upper bounds for global Dantzig.} Let $\rho$ be the height of the condensation graph obtained by contracting each SCC to a vertex, measured by the maximum number of vertices on a directed path. Under the preceding one-internal-action condition on Min, global Dantzig is strongly polynomial whenever $\rho$ is fixed (Theorem~\ref{thm:component-height}). Even without that restriction on Min, if SCC size and the number of actions per state are bounded, any exact-response rule that switches one Max state to a strictly improving action per step makes at most $N2^{O(\rho)}$ switches (Theorem~\ref{thm:bounded-components}). Consequently, $\rho=O(\log N)$ suffices for global Dantzig to be strongly polynomial. The family in Theorem~\ref{thm:small-scc} has $\rho=\Theta(N)$ with bounded SCC size and action degree, showing that exponential dependence on $\rho$ is unavoidable.

\subsection{Related work}

For each fixed discount factor, \citet{ye2011simplex} establishes strong polynomiality (in $N$ and $M$, independent of reward bit-sizes) for Dantzig's rule in stochastic MDPs, and \citet{hansen2013strategy} obtain a corresponding fixed-discount bound for strategy iteration in two-player stochastic games. For single-switch Dantzig and modified simplex in two-player games, the horizon-dependent bounds of \citet{jia2020towards} scale as $O(MNh\log(2Nh))$. When transitions are deterministic, Dantzig achieves a discount-independent bound in MDPs \citep{post2015simplex,hansen2014dantzig} and in forward games, whose directed cycles each belong to one player \citep{mei2026simple}. Beyond those settings, Section~\ref{sec:response-occupation} sharpens the logarithmic factor in the horizon-dependent bound from $\log(2Nh)$ to $\log(N+1)$ whenever every visited Max strategy admits a best response under which all Max states with multiple actions are transient.

\citet{asadi2024deterministic} obtain a deterministic subexponential algorithm for discounted games with unary integer weights by analyzing simultaneous Bellman-greedy strategy updates, a bound that does not directly transfer to single-state Dantzig switches. In the lower-bound instances studied here, rewards can be normalized to $[0,1]$, with binary precision that may grow with the instance size.

Even in single-player MDPs, the choice of improvement rule strongly affects worst-case behavior \citep{melekopoglou1994complexity}. In deterministic discounted MDPs, Howard's simultaneous updates can require exponentially many iterations \citep{zhong2026policy}, although subexponential bounds hold for fixed reward bit-size \citep{mukherjee2025howard}. Improvement rules also differ in how candidate switches are scored: whereas the discounted max-gain bounds of \citet{goenka2026upper} evaluate the largest current one-step advantage within each updated state, modified simplex compares single-state switches by their total gain across all states after the opponent reoptimizes.

Exponential lower bounds are known for several other pivoting and strategy-improvement schemes. \citet{avis2017exponential} establish an exponential lower bound for Cunningham's cyclic action-order rule, and \citet{friedmann2011exponential} proves exponential bounds for locally greedy simultaneous improvement and globally optimizing strategy improvement, the latter searching a current nonworsening action subgraph. Symmetric strategy improvement uses counterstrategies to restrict profitable actions \citep{schewe2015symmetric}, and \citet{van2023worst} establish exponential lower bounds for both that rule and its generalized variant. Neither counterstrategy restrictions nor linear-complementarity pivoting methods such as Lemke's algorithm and the Cottle--Dantzig algorithm \citep{fearnley2010linear} match the single-state exact-response updates studied here.

In stochastic total-reward MDPs, \citet{disser2023unified} give a common exponential lower bound for Dantzig, Bland, and Largest Increase, as well as combinations of these rules, where Largest Increase selects the single switch maximizing the sum of value increases at agent-controlled states. Their return gadgets rely on transition probabilities and auxiliary switches to control reduced costs, an approach also used by \citet{disser2025lower}. The games studied here instead use deterministic transitions and a common discount factor: Min's choice at each return gate caps the current advantage, while modified simplex evaluates each candidate switch against a freshly reoptimized best response.

Both the construction in Theorem~\ref{thm:small-scc} and the discounted counter of \citet{mei2026simplestrategyiterationdantzigsrule} force exponentially many Dantzig switches with at most two actions per state and three states per SCC, and Proposition~\ref{prop:game-source-gates} identifies a local gating condition that recovers the switch order in Mei's unrestricted-cost counter. Expressed in terms of the discount horizon $h$ on $k$ levels, Mei's explicit choice $h=100\cdot 8^{k-2}$ gives $2^k-1=\Theta(h^{1/3})$ switches, whereas the threshold $h=\bar p_k=\Theta(k2^k)$ in Corollary~\ref{cor:small-scc-progress} yields $\Theta(h/\log h)$ switches even to halve the total optimality gap. Mei also proves a superpolynomial discounted lower bound for costs in $\{-1,0,1\}$, along with an exponential lower bound for unrestricted costs and a strongly polynomial implementation for $\{-1,0,1\}$ costs under a mean-payoff rule. In the discounted setting, Theorem~\ref{thm:transfer} provides a general transfer from finite Bland prefixes to Dantzig trajectories, and the two-replica construction forces exponentially many switches even when the algorithm chooses adaptively between Dantzig and modified simplex at each iteration (Theorem~\ref{thm:common-lower-bound}).

For simple stochastic games with a reachability objective, \citet{auger2014almost} call a game \emph{Min-acyclic} if at most one outgoing edge of each Min state lies on a directed cycle. Their solver processes SCCs from downstream to upstream and permanently discards an exit once continuing is strictly better, and \citet{mei2026simple} use the same component order for deterministic forward games. For deterministic discounted games, Theorem~\ref{thm:component-solver} requires the one-internal-edge condition only in mixed SCCs, combining permanent exit elimination with the deterministic-MDP solver of \citet{madani2010discounted} to achieve an $O(N^2M)$ arithmetic bound independent of the discount.

\section{Game model and strategy iteration}\label{sec:model}

\subsection{Game model and basic notation}

A deterministic two-player zero-sum discounted game consists of a finite state set $X=X_{\Max}\mathbin{\dot\cup}X_{\Min}$ partitioned between Max and Min, a nonempty finite action set $\A(x)$ at each state $x\in X$, a successor function $f$, an immediate rational reward $c(a)$ to Max for each action $a$, and a common rational discount factor $\gamma\in(0,1)$. An action $a\in\A(x)$ has source $\src(a)=x$ and unique successor $f(a)\in X$, and distinct actions may share the same successor. Along an infinite play $\omega=(x_0,a_0,x_1,a_1,\ldots)$, the discounted payoff to Max is
\begin{align}\label{eq:payoff}
    \Pi(\omega)=\sum_{j=0}^{\infty}\gamma^j c(a_j).
\end{align}
Because the state and action sets are finite, this series converges absolutely along every play. Write $N=|X|$ and $M=\sum_{x\in X}|\A(x)|$ for the numbers of states and actions.

A stationary deterministic Max strategy $\sigma$ selects one action at every state in $X_{\Max}$, and a stationary deterministic Min strategy $\tau$ selects one action at every state in $X_{\Min}$. Fixing a pair $(\sigma,\tau)$ determines a unique play from each state and an associated value vector $V^{\sigma,\tau}\in\R^X$. With $\sigma$ fixed, let $W^\sigma(x)$ be the infimum of the expected payoff $\mathbb{E}[\Pi(\omega)]$ from $x$ over all Min strategies, including randomized and history-dependent ones. Thus $W^\sigma$ gives the exact-response value when Min optimizes from every state. A stationary Min strategy is an \emph{all-state best response} to $\sigma$ if it attains $W^\sigma(x)$ simultaneously for every $x\in X$.

All rewards and the common discount factor are rational numbers encoded in binary. The discount $\gamma$ may vary with the input, with effective discount horizon $h=(1-\gamma)^{-1}$.

For a candidate vector $z\in\R^X$, define the Bellman operator for a fixed Max strategy $\sigma$ by
\begin{align}\label{eq:bellman}
\begin{alignedat}{2}
    (\mathcal T_\sigma z)(x) &=c(\sigma(x))+\gamma z(f(\sigma(x))), &\qquad &x\in X_{\Max},\\
    (\mathcal T_\sigma z)(x) &=\min_{a\in\A(x)}\{c(a)+\gamma z(f(a))\}, &\qquad &x\in X_{\Min}.
\end{alignedat}
\end{align}
Minimization at $X_{\Min}$ makes $\mathcal T_\sigma$ generally nonlinear, but it remains monotone and a $\gamma$-contraction in the sup norm.

Write $W^*\in\R^X$ for the optimal game value vector. It is the unique fixed point of the game Bellman operator $\mathcal T$, which takes $\max_{a\in\A(x)}\{c(a)+\gamma z(f(a))\}$ at $x\in X_{\Max}$ and $\min_{a\in\A(x)}\{c(a)+\gamma z(f(a))\}$ at $x\in X_{\Min}$. Like $\mathcal T_\sigma$, the operator $\mathcal T$ is a monotone $\gamma$-contraction and satisfies $\mathcal T W^\sigma\ge\mathcal T_\sigma W^\sigma=W^\sigma$. Its iterates from $W^\sigma$ therefore converge monotonically to $W^*\ge W^\sigma$ for every stationary Max strategy $\sigma$.

The full action graph has vertex set $X$ and one directed edge $(x,f(a))$ for every available action $a\in\A(x)$. An SCC is a strongly connected component of this graph, and is called \emph{mixed} if it contains states of both players. An action is \emph{internal} if its source and successor lie in the same SCC, and is an \emph{exit} otherwise. Contracting each SCC to a single vertex and discarding internal edges produces the acyclic condensation graph, whose height $\rho$ is the maximum number of vertices on a directed path.

\subsection{Improvement rules and complexity measures}

At a Max strategy $\sigma$, the current advantage of a Max action $a:u\to w$ is
\begin{align}\label{eq:gameadv}
 G_\sigma(a)=c(a)+\gamma W^\sigma(w)-W^\sigma(u).
\end{align}
Define the switched strategy $\sigma^a$ by $\sigma^a(u)=a$ and $\sigma^a(x)=\sigma(x)$ for $x\ne u$. Assigning unit weight to every Max and Min state yields the total value score $F(\sigma)=\sum_{x\in X}W^\sigma(x)$, under which the exact gain from switching to $a$ at $u$ is
\begin{align}\label{eq:modified-gain}
    F(\sigma^a)-F(\sigma)
      =\sum_{x\in X}\bigl(W^{\sigma^a}(x)-W^\sigma(x)\bigr).
\end{align}
The evaluation subroutine $\operatorname{Evaluate}(\sigma)$ fixes $\sigma$ and solves Min's deterministic MDP for $W^\sigma$, the unique fixed point of~\eqref{eq:bellman}.

An \emph{outer switch} occurs whenever the algorithm commits to changing the action at one Max state, and $T$ denotes the total number of accepted outer switches. All switch lower bounds below hold even if each policy evaluation is charged unit cost. By contrast, a \emph{strongly polynomial} algorithm performs a number of arithmetic operations bounded by a polynomial in $N$ and $M$, independently of the numerical rewards and discount factor, while keeping all intermediate rational numbers of encoding length polynomial in the input size.

Dantzig strategy iteration selects a Max action of largest positive current advantage~\eqref{eq:gameadv}, breaking ties arbitrarily, and terminates with the current strategy when no positive-advantage action exists. The term \emph{global Dantzig} distinguishes this full-game rule from local Dantzig updates inside a single component with fixed boundary values. The modified simplex strategy iteration of \citet{jia2020towards}, abbreviated as \emph{modified simplex}, instead selects a single-state switch maximizing the positive exact gain~\eqref{eq:modified-gain}. In each outer iteration, either rule updates the action at a single Max state and then recomputes an exact all-state Min best response. More generally, for an integer $d\ge1$, depth-$d$ lookahead strategy iteration searches sequences of up to $d$ hypothetical improving switches, evaluating every candidate strategy in the full infinite-horizon discounted game, and executes only the first switch of an optimal sequence. Section~\ref{sec:lookahead-rule} specifies the endpoint score and tie-breaking rule.

\subsection{Bellman properties and component structure}

The next two lemmas characterize best responses to a fixed Max strategy and compare exact-response values before and after a single switch.

\begin{lemma}[Bellman characterization of all-state best responses]\label{lem:bellman}
If $W\in\R^X$ satisfies $\mathcal T_\sigma W=W$, then $W=W^\sigma$. Selecting a minimizing action in~\eqref{eq:bellman} at every Min state yields an all-state stationary best response. Moreover, if every such minimizing action is unique, this strategy is the unique all-state stationary best response, even among randomized stationary strategies.
\end{lemma}
\begin{proof}
Contraction of $\mathcal T_\sigma$ ensures a unique fixed point. Along every play consistent with $\sigma$, iterating the Bellman inequalities gives
\begin{align}\label{eq:bellman-telescope}
W(x_0)\le\sum_{j=0}^{L-1}\gamma^j c(a_j)+\gamma^L W(x_L),
\end{align}
where the remainder $\gamma^L W(x_L)$ tends to zero as $L\to\infty$. A stationary selection of Bellman-minimizing actions attains equality from every initial state, proving $W=W^\sigma$. Finally, any stationary response that assigns positive probability to a strictly suboptimal Bellman action violates its own evaluation equation at that state, establishing uniqueness.
\end{proof}

\begin{lemma}[Value comparison after a single-state switch]\label{lem:strict}\label{lem:gain-sign}
Let $u=\src(a)$, $\sigma'=\sigma^a$, and $g=G_\sigma(a)$. If $g>0$, then
\begin{align}\label{eq:strict}
    W^{\sigma'}\ge W^\sigma, \qquad W^{\sigma'}(u)\ge W^\sigma(u)+g>W^\sigma(u).
\end{align}
If $g<0$, then $W^{\sigma'}\le W^\sigma$ with strict inequality at $u$. If $g=0$, then $W^{\sigma'}=W^\sigma$. Consequently, the exact all-state gain $F(\sigma')-F(\sigma)$ has the same sign as the current advantage $g$.
\end{lemma}
\begin{proof}
Write $v=W^\sigma$ and let $\mathbf e_u$ be the unit coordinate vector at $u$. Then $\mathcal T_{\sigma'}v=v+g \mathbf e_u$, and iterating the monotone contraction $\mathcal T_{\sigma'}$ from $v$ converges to $W^{\sigma'}$. Monotonicity yields $W^{\sigma'}\ge v+g \mathbf e_u\ge v$ when $g>0$, proving~\eqref{eq:strict}, and $W^{\sigma'}\le v+g \mathbf e_u\le v$ with strict inequality at $u$ when $g<0$. When $g=0$, $v$ is already the fixed point of $\mathcal T_{\sigma'}$. Summing the coordinate differences gives the sign assertion.
\end{proof}

By Lemma~\ref{lem:strict}, a sequence of positive-advantage switches never revisits a Max strategy, and when no such switch exists, the current strategy satisfies $\mathcal T W^\sigma=W^\sigma$ and is optimal. Negating all rewards converts the largest positive advantage into the most negative reduced cost, matching the minimization convention of \citet{mei2026simple}.

\begin{definition}\label{def:one-internal}
Let $\mathcal C_1$ be the class of deterministic discounted games in which, within every mixed SCC, each Min state has at most one internal action. No restriction is placed on SCCs controlled by a single player.
\end{definition}

Mixed SCCs in $\mathcal C_1$ may have arbitrary size and may contain directed cycles under stationary strategy pairs. Section~\ref{sec:structure} shows that every game in $\mathcal C_1$ can be solved in $O(N^2M)$ arithmetic operations. Constructing the lower-bound families inside $\mathcal C_1$ thus separates the trajectory length of global strategy iteration from the arithmetic cost of solving these games directly.

\section{Exponential lower bounds for Dantzig and modified simplex}\label{sec:main-results}

The two main lower bounds come from two families of games in $\mathcal C_1$. Sections~\ref{sec:transfer} and~\ref{sec:lookahead} construct these families from the same acyclic binary counter. In the first family, exactly one action attains the largest positive advantage at each Dantzig switch. In the second, duplicating each state of the counter forces every trajectory, whether each step uses Dantzig's rule or modified simplex, to execute each counter switch across both copies before advancing.

\begin{theorem}[Exponential lower bound for Dantzig's rule]\label{thm:small-scc}
For every $k\ge1$, there is a game in $\mathcal C_1$ with at most two actions per state and at most three states per SCC such that, from a specified initial Max strategy, Dantzig makes exactly $T$ switches, where
\begin{align}\label{eq:small-scc-bound}
    N=6k+6,\qquad M=12k+8,\qquad T=2^{k+1}-k-2.
\end{align}
Dantzig's choice is unique at every switch, and every stationary deterministic Max strategy admits a unique all-state stationary Min best response. The game has nonnegative rewards and an $O(k^2)$-bit encoding.
\end{theorem}

\begin{theorem}[A common exponential lower bound for Dantzig and modified simplex]\label{thm:common-lower-bound}
For every $k\ge1$, there is a game in $\mathcal C_1$ with $N=12k+12$ states and an initial Max strategy from which every run that chooses at each step between Dantzig's rule and modified simplex makes exactly $2(2^{k+1}-k-2)$ switches. The choice of rule at each iteration may depend on the entire history, and ties may be broken arbitrarily.
\end{theorem}

Section~\ref{sec:dantzig-lower} proves Theorem~\ref{thm:small-scc} via the Bland-to-Dantzig transfer of Section~\ref{sec:transfer}, and Section~\ref{sec:cyclic-construction} proves Theorem~\ref{thm:common-lower-bound} using a two-replica cyclic gate gadget. The remaining results of this section quantify optimality-gap reduction and direct solution complexity on the first family.

Let $\widehat\sigma_t$ be the Max strategy after $t$ Dantzig switches and $e_t(x)=W^*(x)-W^{\widehat\sigma_t}(x)$ be the optimality gap at state $x$. For $e_0(x)>0$ and $0<\theta<1$, let $T_\theta(x)$ be the first iteration $t$ with $e_t(x)\le\theta e_0(x)$. The total optimality gap and its halving time are
\begin{equation*}
    E_t=\sum_{x\in X}e_t(x),\qquad
    T^{\mathrm{sum}}_{1/2}=\min\{t\in\{0,\ldots,T\}:E_t\le E_0/2\}.
\end{equation*}

In the construction of Theorem~\ref{thm:small-scc} (Section~\ref{sec:transfer}), the action rewards are parameterized by the rational discount $\gamma$, whereas the game graph, uniqueness properties, and Dantzig trajectory remain unchanged whenever
\begin{align}\label{eq:small-scc-horizon}
    h=(1-\gamma)^{-1}\ge \bar p_k:=64(k+1)(12\cdot2^k+1).
\end{align}

\begin{corollary}[Gap reduction and discount-horizon lower bounds]\label{cor:small-scc-progress}
Under~\eqref{eq:small-scc-horizon}, let $t_*:=2^k-k$. For the $k$-level game of Theorem~\ref{thm:small-scc}, the total gap and the gap at a designated Max state $\xi_k$ (at level $k$ of the counter in Section~\ref{sec:acyclic-source}) satisfy
\begin{align}
    T^{\mathrm{sum}}_{1/2}&\ge t_*,\label{eq:small-scc-total-half}\\
    T_\theta(\xi_k)&=t_*,\qquad 0<\theta<1.\label{eq:small-scc-half}
\end{align}
Specifically, $e_t(\xi_k)$ remains at its positive initial value until switch $t_*$, when it vanishes. Consequently, for every rational $h\ge\bar p_1$, this family contains a game with $N,M=\Theta(\log h)$ in which, for all $0<\theta<1$,
\begin{equation}\label{eq:small-scc-horizon-bound}
    T^{\mathrm{sum}}_{1/2}=\Omega(h/\log h),\qquad T_\theta(\xi_k)=\Omega(h/\log h).
\end{equation}
\end{corollary}

Section~\ref{app:progress-proof} proves Corollary~\ref{cor:small-scc-progress}. Because $E_t/N$ is the expected optimality gap under a uniform initial distribution, the same halving lower bound applies to the average gap. For comparison, the one-step gap contraction proved in Section~\ref{app:response-occupation-proof} (see~\eqref{eq:occupation-contraction}) gives $T^{\mathrm{sum}}_{1/2}\le\lceil Nh\log 2\rceil$ for both Dantzig and modified simplex without any state-visitation assumption. At $N,M=\Theta(\log h)$, the Dantzig lower bound $\Omega(h/\log h)$ and this general upper bound $O(h\log h)$ differ by at most a factor of $\Theta((\log h)^2)$.

The explicit discount $\gamma=1-1/\bar p_k$ in Theorem~\ref{thm:small-scc} gives a total binary encoding length of $O(k^2)$, whereas for an arbitrary admissible discount of bit length $b_\gamma$, the encoding length is $O(k(k+b_\gamma))$.

\begin{corollary}[Linear arithmetic complexity of the Dantzig instances]\label{cor:small-scc-linear}
For the family in Theorem~\ref{thm:small-scc}, optimal values at all states and a stationary optimal strategy pair can be computed in $O(N+M)$ arithmetic operations.
\end{corollary}
\begin{proof}
Process the SCCs in reverse topological order, replacing each exit $a:x\to y$ by a terminal action with fixed payoff $c(a)+\gamma W^*(y)$. Because each SCC has at most three states and at most two actions per state, it admits at most eight local stationary strategy pairs, so evaluating all of them and selecting one that satisfies the local Bellman equations takes $O(1)$ arithmetic operations. Both SCC decomposition and the sum of local solves therefore take $O(N+M)$ operations, and backward induction guarantees global optimality of the assembled strategies. Finally, every downstream value arises from a stationary strategy pair on at most $N$ states and hence has polynomial bit length, and each local solve combines only constantly many such values and input coefficients, keeping all intermediate rational numbers of polynomial bit length.
\end{proof}

Because bounded-size components allow trivial local enumeration, the exponential switch count in Theorem~\ref{thm:small-scc} reflects the myopia of global Dantzig updates rather than any intrinsic difficulty in solving these games.

\section{From Bland to Dantzig via return gates}\label{sec:transfer}

The return-gate construction embeds any finite sequence of Bland switches from a single-player MDP into a discounted two-player game in which Dantzig follows the same switch sequence. Applying this transfer to an acyclic binary counter proves Theorem~\ref{thm:small-scc}. Comparing the resulting game with the forward-only variant obtained by deleting Min's return actions isolates how opponent responses drive the switch count, and a general gain identity quantifies the value lost when Min reoptimizes.

\subsection{The return-gate mechanism}\label{sec:gate-mechanism}

A return gate lets Min send play back to an action's starting state, capping its current advantage so that Dantzig reproduces Bland's switch order.

The source is a deterministic MDP that maximizes undiscounted total reward. It has $n$ states in $S$, $m$ actions, and an absorbing sink $s$ with a single zero-reward self-loop. Each action $a:u\to w$ has a rational reward $r_a$ and a distinct rank $\rank(a)\in\{1,\ldots,m\}$. Bland's rule switches to the lowest-ranked action with positive undiscounted advantage and stops when no such action exists.

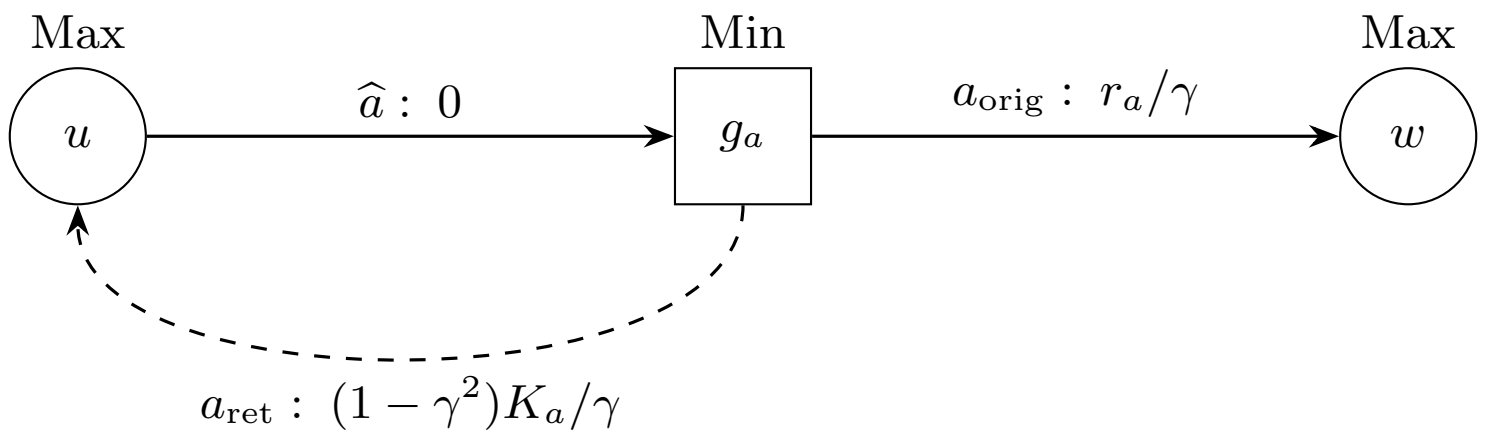
\begin{figure}[htbp]
\centering
\begin{tikzpicture}[font=\small]
\node[state,label=above:{$\Max$}] (u) at (0,0) {$u$};
\node[gate,label=above:{$\Min$}] (g) at (4.4,0) {$g_a$};
\node[state,label=above:{$\Max$}] (w) at (8.8,0) {$w$};
\draw[edge] (u) -- node[above] {$\widehat a:\ 0$} (g);
\draw[edge] (g) -- node[above] {$a_{\orig}:\ r_a/\gamma$} (w);
\draw[edge,dashed] (g.south) .. controls (4.4,-1.8) and (0,-1.8) ..
  node[midway,below] {$a_{\ret}:\ (1-\gamma^2) K_a/\gamma$} (u.south);
\end{tikzpicture}
\caption{Return gate for a source action $a:u\to w$. Edges show actions and immediate rewards. The dashed edge returns to $u$. If $u=w$, both circles denote the same state.}\label{fig:gate}
\end{figure}

In the target game, Max controls all original states, and each source action $a:u\to w$ is replaced by a zero-reward Max action $\widehat a:u\to g_a$ leading to a Min gate $g_a$ (Figure~\ref{fig:gate}). At $g_a$, Min can either go forward to $w$ for reward $r_a/\gamma$ or return to $u$ for reward $(1-\gamma^2)K_a/\gamma$. Because $\widehat a$ pays zero and discounts the gate's reward by $\gamma$, these two choices yield two-step discounted rewards $r_a$ and $(1-\gamma^2)K_a$ as seen from $u$. In particular, $K_a$ is the infinite-horizon payoff from $u$ if Max repeatedly plays $\widehat a$ and Min always returns. Section~\ref{sec:gate-parameters} sets $K_a$ as a decreasing function of $\rank(a)$. For any source policy $\sigma$, the lifted Max strategy $\widehat\sigma$ selects $\widehat a$ wherever $\sigma$ selects $a$.

Fix a proper source policy $\sigma$, meaning one that reaches $s$ from every state in finitely many steps. Let $V_\sigma^\lambda$ denote its source value vector at discount $\lambda$, so $V_\sigma^1$ records the undiscounted total reward up to absorption. The source advantage of $a:u\to w$ is
\begin{align}\label{eq:sourceadv}
    A_\sigma^\lambda(a)=r_a+\lambda V_\sigma^\lambda(w)-V_\sigma^\lambda(u),
    \qquad \lambda\in\{1,\gamma^2\}.
\end{align}
Bland's rule uses $A_\sigma^1$. Because traversing a gate forward takes two transitions in the target game, the effective discount per source step is $\gamma^2$. Abbreviate $V_\sigma=V_\sigma^{\gamma^2}$ and $A_\sigma=A_\sigma^{\gamma^2}$.

Under the parameter conditions in Section~\ref{sec:gate-parameters}, every original state satisfies $W^{\widehat\sigma}(u)=V_\sigma(u)$. Discounting the two gate branches back to $u$ and subtracting $V_\sigma(u)$ gives the clipping identity
\begin{equation}\label{eq:clip}
    G_{\widehat\sigma}(\widehat a)=\min\bigl\{A_\sigma(a),\,(1-\gamma^2)(K_a-V_\sigma(u))\bigr\}.
\end{equation}
Write $C_\sigma(a)=(1-\gamma^2)(K_a-V_\sigma(u))$ for the second term, which is the advantage of returning to $u$. Because Min is free to choose the return branch at $g_a$, $C_\sigma(a)$ caps the current advantage of $\widehat a$ (Lemma~\ref{lem:preserve}).

Along the prescribed Bland prefix, every source action with $A_\sigma^1(a)>0$ satisfies $A_\sigma(a)>C_\sigma(a)$, so Min chooses the return branch at $g_a$ and $\widehat a$ attains its cap $C_\sigma(a)$. Among these improving actions, the one of smallest Bland rank has the largest cap, whereas every action with $A_\sigma^1(a)\le0$ has target advantage strictly below all caps (Lemma~\ref{lem:order}). Once Max switches to $\widehat a$, the updated source policy $\sigma'$ still satisfies $V_{\sigma'}(u)<K_a$, causing Min's new best response to switch to the forward branch at $g_a$. Proposition~\ref{prop:response-mechanism} quantifies how this response change reduces the realized gain.

\subsection{Absorbing source prefixes and perturbation bounds}\label{sec:perturb}

Because each gate's forward branch evaluates the discounted advantage $A_\sigma(a)$ rather than $A_\sigma^1(a)$, the discount factor must be close enough to one to separate positive undiscounted advantages from nonpositive ones.

Fix a finite Bland prefix $\sigma_0,\ldots,\sigma_T$ consisting of proper policies, and let $H\ge1$ be an integer such that each visited policy reaches $s$ within $H$ transitions from every state. One may always take $H=\max\{1,n-1\}$, since no state repeats along a simple path to $s$. Because the source rewards are rational, there exist integers $R,D\ge1$ satisfying $|r_a|\le R$ and $Dr_a\in\Z$ for every action $a$.

\begin{lemma}[Discount perturbation and advantage gap]\label{lem:perturb}
For each policy $\sigma$ on the prefix,
\begin{align}
    \max\{\norm{V_\sigma},\norm{V_\sigma^1}\}&\le HR,\label{eq:boundB}\\
    \norm{V_\sigma-V_\sigma^1}&\le(1-\gamma^2)R\binom{H}{2},\label{eq:Verror}\\
    \norm{A_\sigma-A_\sigma^1}&\le(1-\gamma^2)H^2R.\label{eq:Aerror}
\end{align}
Every undiscounted advantage lies in $D^{-1}\Z$, so every positive undiscounted advantage is at least $1/D$.
\end{lemma}
\begin{proof}
Each policy path to $s$ collects at most $H$ rewards of magnitude at most $R$, proving~\eqref{eq:boundB}. Summing the weight differences $1-\gamma^{2j}\le j(1-\gamma^2)$ over $0\le j<H$ yields~\eqref{eq:Verror}. Subtracting the two advantage expressions in~\eqref{eq:sourceadv} gives
\[
    \norm{A_\sigma-A_\sigma^1}
    \le (1+\gamma^2)\norm{V_\sigma-V_\sigma^1}
       +(1-\gamma^2)\norm{V_\sigma^1},
\]
and combining~\eqref{eq:boundB} and~\eqref{eq:Verror} with $2\binom{H}{2}+H=H^2$ proves~\eqref{eq:Aerror}. Finally, every undiscounted value is a finite sum of rewards in $D^{-1}\Z$, so~\eqref{eq:sourceadv} places every undiscounted advantage in $D^{-1}\Z$.
\end{proof}

\subsection{Choosing gate parameters and the discount}\label{sec:gate-parameters}

The gate parameters are set so that the return caps decrease strictly with Bland rank, and $\gamma$ is chosen close enough to one to place all caps strictly between the forward advantages of nonimproving and improving source actions.

Before iteration, fix nonnegative rational bounds $B_V$ and $B_A$, independent of the prefix length $T$, such that $\norm{V_\sigma}\le B_V$ and $\norm{A_\sigma-A_\sigma^1}\le(1-\gamma^2)B_A$ for every visited policy $\sigma$ and every $\gamma\in(0,1)$. Lemma~\ref{lem:perturb} permits the default choice $(B_V,B_A)=(HR,H^2R)$. Set
\begin{align}
    K_a &= B_V+B_A+1+(m-\rank(a))(2B_V+1),\label{eq:K}\\
    Q &= K_{\max}+B_V+B_A+1, \qquad K_{\max}=\max_a K_a,\label{eq:Q}\\
    p &= 16DQ, \qquad h\ge p.\label{eq:discount}
\end{align}
Here the offset $B_V+B_A+1$ keeps every cap above the discount error $(1-\gamma^2)B_A$, while the rank spacing $2B_V+1$ exceeds the maximum gap $2B_V$ between any two source values. Any rational discount $\gamma\in(0,1)$ with horizon $h\ge p$ is admissible, including $\gamma=1-1/p$, and satisfies
\begin{equation}\label{eq:deltabound}
    1-\gamma^2=\frac{1+\gamma}{h}<\frac{2}{h}\le\frac{1}{8DQ}.
\end{equation}

Replacing every source action, including the sink self-loop, by the return gate of Section~\ref{sec:gate-mechanism} produces a deterministic two-player game with $n+m$ states and $3m$ actions in which control alternates on every transition.

Taking $H=\max\{1,n-1\}$, the smallest admissible integer $R$, and the product of the reward denominators for $D$ keeps all three integers of polynomial bit length. Consequently, the default bounds together with $\gamma=1-1/p$ produce a target game of encoding length polynomial in the source input size, independently of $T$. For any supplied admissible rational discount, the encoding length is polynomial in the combined bit length of the source and $\gamma$.

\subsection{Preservation of policy values and clipping of advantages}

At the gate of any currently selected action, the forward branch preserves the source policy value at the starting state whereas returning yields a strictly larger payoff, so Min uniquely goes forward.

\begin{lemma}[Values and advantages under the lifted strategy]\label{lem:preserve}
For any source policy $\sigma$ with $\norm{V_\sigma}\le B_V$, the exact-response value under $\widehat\sigma$ satisfies
\begin{align}
    W^{\widehat\sigma}(u)&=V_\sigma(u), &&u\in S,\label{eq:Woriginal}\\
    \gamma W^{\widehat\sigma}(g_a)&=V_\sigma(u)+\min\{A_\sigma(a),C_\sigma(a)\}, &&a:u\to w.\label{eq:Wgate}
\end{align}
The forward branch is uniquely minimizing at each selected gate, and the clipping identity~\eqref{eq:clip} holds for every action.
\end{lemma}
\begin{proof}
Take $V_\sigma(u)$ as the candidate value at each original state $u$. Discounted back to $u$, the forward and return branches at $g_a$ evaluate to $V_\sigma(u)+A_\sigma(a)$ and $V_\sigma(u)+C_\sigma(a)$, whose minimum gives~\eqref{eq:Wgate}. When $a=\sigma(u)$, policy evaluation gives $A_\sigma(a)=0$, whereas $K_a>B_V\ge V_\sigma(u)$ gives $C_\sigma(a)>0$. Hence the forward branch is uniquely minimizing at $g_a$, and its discounted value $V_\sigma(u)$ satisfies the Bellman equation at $u$. Lemma~\ref{lem:bellman} therefore yields $W^{\widehat\sigma}$, and subtracting~\eqref{eq:Woriginal} from~\eqref{eq:Wgate} gives~\eqref{eq:clip} via~\eqref{eq:gameadv}.
\end{proof}

\subsection{The Dantzig switch and Min's best response}

Under~\eqref{eq:clip}, every improving source action attains its return cap in the target game, and these caps strictly decrease with Bland rank while remaining above the target advantage of every nonimproving action.

\begin{lemma}[Dantzig's unique choice]\label{lem:order}
If Bland selects $a_*$ at a policy $\sigma$ on the source prefix, then Dantzig uniquely selects $\widehat a_*$ at $\widehat\sigma$, with
\begin{equation}\label{eq:rankgap}
    G_{\widehat\sigma}(\widehat a_*)-G_{\widehat\sigma}(\widehat b)\ge 1-\gamma^2
    \qquad (b\ne a_*).
\end{equation}
\end{lemma}
\begin{proof}
Fix a policy $\sigma$ on the prefix. For each action $b:u\to w$, $|V_\sigma(u)|\le B_V$ together with~\eqref{eq:K} and~\eqref{eq:Q} gives $B_A+1\le K_b-V_\sigma(u)<Q$. Multiplying by $1-\gamma^2$ and using~\eqref{eq:deltabound} yields
\begin{equation}\label{eq:capbounds}
    (1-\gamma^2)(B_A+1)\le C_\sigma(b)<\frac{1}{8D}.
\end{equation}
By Lemma~\ref{lem:perturb}, every positive undiscounted advantage is at least $1/D$, whereas the discount perturbation is at most $(1-\gamma^2)B_A<1/(8D)$ by~\eqref{eq:deltabound} and $B_A<Q$. Thus, for any actions $a$ and $b$,
\begin{equation}\label{eq:nonpositive}
\begin{aligned}
    A_\sigma^1(a)>0  &\implies A_\sigma(a)-C_\sigma(b)>3/(4D),\\
    A_\sigma^1(a)\le0 &\implies A_\sigma(a)-C_\sigma(b)\le -(1-\gamma^2).
\end{aligned}
\end{equation}
By~\eqref{eq:clip}, every improving source action attains its positive cap, while every nonimproving action has target advantage at least $1-\gamma^2$ below every cap.

For any two improving actions $a$ and $b$ with $\rank(a)<\rank(b)$,~\eqref{eq:K} gives $K_a-K_b\ge 2B_V+1$. Because the values at their starting states differ by at most $2B_V$, it follows that $C_\sigma(a)-C_\sigma(b)\ge(1-\gamma^2)[(2B_V+1)-2B_V]=1-\gamma^2$. Since Bland selects the improving action of smallest rank,~\eqref{eq:rankgap} follows.
\end{proof}

\begin{lemma}[Uniqueness of all-state stationary Min best responses]\label{lem:unique}
For each policy $\sigma$ on the source prefix, including its endpoint $\sigma_T$, the lifted strategy $\widehat\sigma$ admits a unique all-state stationary Min best response. At $g_a$, this response takes the return branch if $A_\sigma^1(a)>0$ and the forward branch if $A_\sigma^1(a)\le0$, and the two Bellman branch values differ in magnitude by at least $(1-\gamma^2)/\gamma$.
\end{lemma}
\begin{proof}
By Lemma~\ref{lem:preserve}, the forward branch value minus the return branch value at $g_a$ equals
\begin{equation}\label{eq:branchdiff}
    \frac{A_\sigma(a)-C_\sigma(a)}{\gamma}.
\end{equation}
Setting $b=a$ in~\eqref{eq:nonpositive} shows that the numerator exceeds $3/(4D)>1-\gamma^2$ when $A_\sigma^1(a)>0$ and is at most $-(1-\gamma^2)$ otherwise. Its sign therefore determines Min's unique optimal branch at every gate, and Lemma~\ref{lem:bellman} extends uniqueness to all stationary responses, including randomized ones.
\end{proof}

\subsection{The general transfer theorem}

\begin{theorem}[Exact transfer from Bland to Dantzig]\label{thm:transfer}
Let $\sigma_0,\ldots,\sigma_T$ be a finite Bland prefix of proper source policies, and form the target game of Section~\ref{sec:gate-mechanism} using any gate parameters and rational discount admissible in Section~\ref{sec:gate-parameters}. Starting from the lifted strategy $\widehat\sigma_0$ under exact Min best responses, whenever Bland switches from $\sigma_t$ to $\sigma_{t+1}$ ($0\le t<T$) by selecting $a$, Dantzig at $\widehat\sigma_t$ uniquely selects $\widehat a$ and transitions to $\widehat\sigma_{t+1}$. Moreover, Min has a unique all-state stationary best response to every $\widehat\sigma_t$ for $0\le t\le T$.
\end{theorem}
\begin{proof}
Proceed by induction on $t$ from $\widehat\sigma_0$. For $0\le t<T$, Lemma~\ref{lem:order} shows that Dantzig at $\widehat\sigma_t$ uniquely selects the lift of Bland's chosen action, yielding $\widehat\sigma_{t+1}$ by definition, and Lemma~\ref{lem:unique} gives uniqueness of the all-state stationary Min best response for all $0\le t\le T$.
\end{proof}

\begin{remark}
Each Bland switch increases the undiscounted value at its state by at least $1/D$ and decreases no other state's value. Summing over the $n$ states, whose values lie in $[-B_V,B_V]$, gives $T\le 2nB_VD$. Since $m\ge n$,~\eqref{eq:K}--\eqref{eq:discount} give $Q=2mB_V+2B_A+m+1>2nB_V$ and $h\ge16DQ$, so $T<h/16$ along the reproduced Bland prefix.

At the endpoint $\sigma_T$, an unselected action with $A_{\sigma_T}^1(a)=0$ can acquire a small positive discounted advantage, allowing Dantzig to continue past $\widehat\sigma_T$. If instead every unselected action at $\sigma_T$ has strictly negative undiscounted advantage, Lemma~\ref{lem:perturb} bounds each such advantage above by $-1/D$, and the perturbation bound $(1-\gamma^2)B_A<1/(8D)$ keeps $A_{\sigma_T}(a)$ negative, so~\eqref{eq:clip} guarantees termination after exactly $T$ switches.

Theorem~\ref{thm:transfer} does not require the source graph to be acyclic, as Section~\ref{sec:source} illustrates on the cyclic Bland counter of \citet{disser2023unified}. When every source policy is proper and satisfies the bounds $B_V$ and $B_A$, Lemmas~\ref{lem:preserve} and~\ref{lem:unique} hold at every Max strategy in the target game, as occurs for the acyclic source below.
\end{remark}

\subsection{The acyclic Bland counter}\label{sec:acyclic-source}

Because every policy of the acyclic $k$-bit counter reaches the sink, Theorem~\ref{thm:transfer} transfers its $2^{k+1}-k-2$ Bland switches to Dantzig on $N=6k+6$ states, establishing a $2^{\Omega(N)}$ lower bound. This counter adapts the layered graph of \citet{disser2025lower} (full version, Figure~5.1 and Lemma~5.4) by removing their random transition perturbations and fixing the reward base to two.

For $k\ge1$, the source MDP $\mathcal M_k$ has two states $\alpha_i,\xi_i$ at each level $i=1,\ldots,k+1$, together with a sink $s$. At each state $u\in\{\alpha_i,\xi_i\}$ of level $i\le k$, the two available actions are
\begin{align}\label{eq:acyclic-source}
    a_\alpha(u) &: u\longrightarrow\alpha_{i+1} &&\text{with reward }0,\notag\\
    a_\xi(u) &: u\longrightarrow\xi_{i+1} &&\text{with reward }2^i.
\end{align}
Write $a_\alpha$ and $a_\xi$ when the starting state is clear from context. At level $k+1$, forced actions lead from $\alpha_{k+1}$ and $\xi_{k+1}$ to $s$ with rewards $2^{k+1}$ and $0$, respectively, and $s$ has a forced zero-reward self-loop. Thus $n=2k+3$, $m=4k+3$, every policy reaches $s$ within $H=k+1$ transitions, and all rewards are integers bounded by $R=2^{k+1}$.

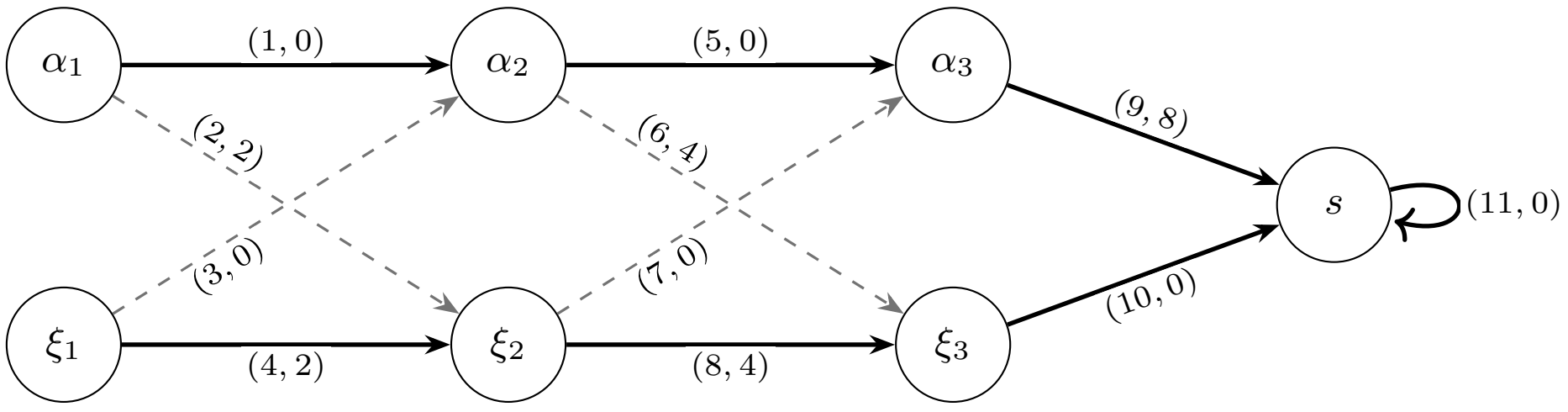
\begin{figure}[htbp]
\centering
\begin{tikzpicture}[font=\small,
  chosen/.style={edge,line width=1pt},
  other/.style={edge,dashed,draw=black!55},
  alabel/.style={font=\scriptsize,fill=white,inner sep=1pt}]
\node[state] (a1) at (0,0) {$\alpha_1$};
\node[state] (b1) at (0,-2.2) {$\xi_1$};
\node[state] (a2) at (3.5,0) {$\alpha_2$};
\node[state] (b2) at (3.5,-2.2) {$\xi_2$};
\node[state] (a3) at (7,0) {$\alpha_3$};
\node[state] (b3) at (7,-2.2) {$\xi_3$};
\node[state] (s) at (10,-1.1) {$s$};
\draw[chosen] (a1) -- node[alabel,above] {$(1,0)$} (a2);
\draw[other]  (a1) -- node[alabel,above,sloped,pos=0.30] {$(2,2)$} (b2);
\draw[other]  (b1) -- node[alabel,below,sloped,pos=0.30] {$(3,0)$} (a2);
\draw[chosen] (b1) -- node[alabel,below] {$(4,2)$} (b2);
\draw[chosen] (a2) -- node[alabel,above] {$(5,0)$} (a3);
\draw[other]  (a2) -- node[alabel,above,sloped,pos=0.30] {$(6,4)$} (b3);
\draw[other]  (b2) -- node[alabel,below,sloped,pos=0.30] {$(7,0)$} (a3);
\draw[chosen] (b2) -- node[alabel,below] {$(8,4)$} (b3);
\draw[chosen] (a3) -- node[alabel,above,sloped] {$(9,8)$} (s);
\draw[chosen] (b3) -- node[alabel,below,sloped] {$(10,0)$} (s);
\draw[chosen] (s) edge[loop right] node[alabel,right] {$(11,0)$} (s);
\end{tikzpicture}
\caption{The two-bit counter $\mathcal M_2$. Each edge label gives the Bland rank followed by the immediate reward. At a two-action state, $a_\alpha$ leads to the upper row and $a_\xi$ to the lower row. Solid edges form the initial policy and dashed edges are the alternatives.}\label{fig:acyclic-example}
\end{figure}

Bland ranks are assigned level by level from $i=1$ up to $k+1$, placing the sink loop last. Within level $i$, the actions at $\alpha_i$ precede those at $\xi_i$, and at each two-action state $a_\alpha$ precedes $a_\xi$. The initial policy stays in its own row, selecting $a_\alpha$ at $\alpha_i$ and $a_\xi$ at $\xi_i$ for all $i\le k$ (Figure~\ref{fig:acyclic-example}).

For $\nu=\sum_{i=1}^k\nu_i2^{i-1}$ with $\nu_i\in\{0,1\}$, a policy is \emph{canonical for $\nu$} if, at every level $i\le k$, $\alpha_i$ selects the action of larger undiscounted value and $\alpha_i,\xi_i$ choose the same destination if and only if $\nu_i=1$. Backward induction determines a unique canonical policy for each $\nu$ (Section~\ref{app:acyclic-counter-proof}), and the initial policy is canonical for $\nu=0$.

\begin{lemma}[Exact switch count for the acyclic counter]\label{lem:acyclic-counter}
On $\mathcal M_k$, Bland's rule makes exactly
\begin{align}\label{eq:acyclic-count}
    T_k=\sum_{j=1}^k j2^{k-j}=2^{k+1}-k-2
\end{align}
strictly improving switches before reaching an optimal policy, visiting the canonical policies for $\nu=0,1,\ldots,2^k-1$ in increasing order.
\end{lemma}

At the canonical policy for $\nu<2^k-1$, the improving actions are precisely those at states $\xi_i$ with $\nu_i=0$. Let $j$ be the lowest index with $\nu_j=0$. Bland first switches $\xi_j$ to match the destination of $\alpha_j$. When $j>1$, $\alpha_j$ and $\xi_j$ now share the same continuation value, so the immediate reward $2^{j-1}$ on $a_\xi$ makes $\xi_j$ the better successor from level $j-1$. Because lower levels have smaller Bland ranks, this triggers a cascade of switches at $\alpha_{j-1},\ldots,\alpha_1$ before any pending switch at levels above $j$. Together, these $j$ switches set bit $j$ to one and reset bits $1,\ldots,j-1$ to zero, reaching the canonical policy for $\nu+1$. Counting the $2^{k-j}$ binary increments whose lowest zero bit is $j$ yields~\eqref{eq:acyclic-count}, as verified in Section~\ref{app:acyclic-counter-proof}.

In Figure~\ref{fig:acyclic-example}, the two initial improving actions lie at $\xi_1$ (rank $3$, undiscounted advantage $2$) and $\xi_2$ (rank $7$, undiscounted advantage $4$). Bland selects rank $3$ first, and with return gates Dantzig reproduces all four switches at $\xi_1,\xi_2,\alpha_1,\xi_1$ in order. Without return branches, Dantzig instead switches $\xi_2$ and then $\alpha_1$ before terminating in two steps (Proposition~\ref{prop:forward-only}).

Along every policy path in $\mathcal M_k$, the undiscounted total reward is at most
\begin{align*}
    \sum_{i=1}^k2^i+2^{k+1}=2R-2<2R.
\end{align*}
Uniformly over all source policies and all $\gamma\in(0,1)$, one may therefore set
\begin{align}\label{eq:acyclic-sharp-bounds}
    B_V=2R,\qquad B_A=(4H+2)R,\qquad H=k+1,\qquad R=2^{k+1}.
\end{align}
Indeed, since each path collects at most $H$ rewards and $1-\gamma^{2j}\le j(1-\gamma^2)$,
\begin{align*}
    |V_\sigma(u)-V_\sigma^1(u)|&\le 2(1-\gamma^2) HR,\\
    |A_\sigma(a)-A_\sigma^1(a)|
       &\le (1+\gamma^2)2(1-\gamma^2) HR+2(1-\gamma^2) R
        \le(1-\gamma^2)(4H+2)R.
\end{align*}
Because~\eqref{eq:acyclic-sharp-bounds} holds for every source policy, every Max strategy in the target game admits a unique all-state stationary Min best response. Acyclicity of $\mathcal M_k$ also confines each nonterminal target SCC to one original state and its outgoing gates (Section~\ref{sec:dantzig-lower}). Section~\ref{sec:lookahead} builds on the same counter.

\subsection{Proof of Theorem~\ref{thm:small-scc}}\label{sec:dantzig-lower}

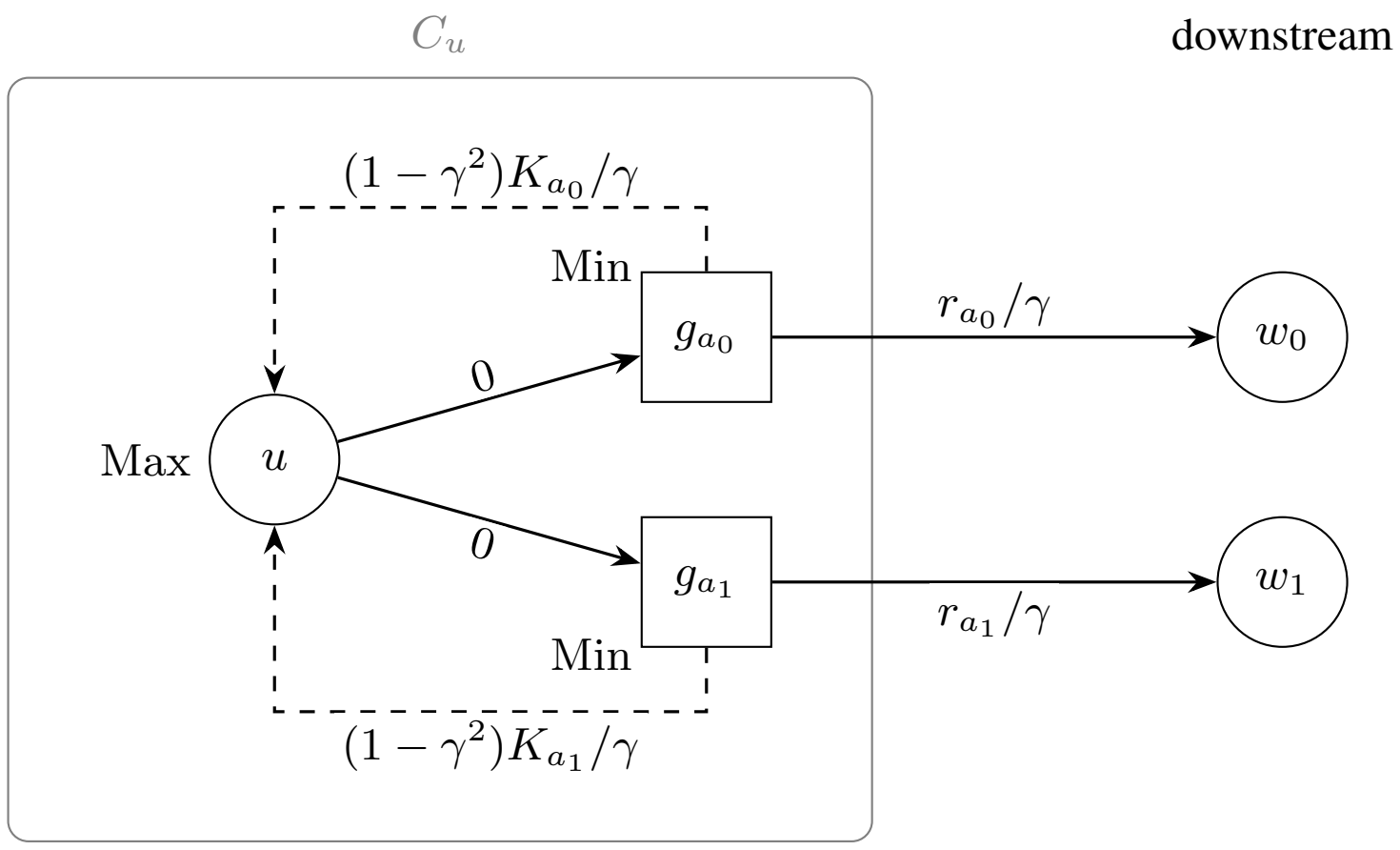
\begin{figure}[htbp]
\centering
\begin{tikzpicture}[font=\small,
  reward/.style={fill=white,inner sep=1.5pt}]
\draw[rounded corners,gray] (-1.85,-2.65) rectangle (4.15,2.65);
\node[gray] at (1.15,2.95) {$C_u$};
\node[state,label=left:{$\Max$}] (u) at (0,0) {$u$};
\node[gate] (g0) at (3,0.85) {$g_{a_0}$};
\node[gate] (g1) at (3,-0.85) {$g_{a_1}$};
\node at (2.2,1.35) {$\Min$};
\node at (2.2,-1.35) {$\Min$};
\node[state] (w0) at (7,0.85) {$w_0$};
\node[state] (w1) at (7,-0.85) {$w_1$};
\node at (7,2.95) {downstream};
\draw[edge] (u) -- node[reward,above,sloped] {$0$} (g0);
\draw[edge] (u) -- node[reward,below,sloped] {$0$} (g1);
\draw[edge,dashed] (g0.north) -- (3,1.75) --
  node[reward,above] {$(1-\gamma^2) K_{a_0}/\gamma$} (0,1.75) -- (u.north);
\draw[edge,dashed] (g1.south) -- (3,-1.75) --
  node[reward,below] {$(1-\gamma^2) K_{a_1}/\gamma$} (0,-1.75) -- (u.south);
\draw[edge] (g0) -- node[reward,above] {$r_{a_0}/\gamma$} (w0);
\draw[edge] (g1) -- node[reward,below] {$r_{a_1}/\gamma$} (w1);
\end{tikzpicture}
\caption{Three-state component $C_u$. At each gate, the dashed return to $u$ is the only internal action. The forward branches leave $C_u$ for $w_0$ and $w_1$, whose values can change after downstream switches.}\label{fig:small-scc}
\end{figure}

\begin{proof}
Starting from $\widehat\sigma_0$, Theorem~\ref{thm:transfer} forces Dantzig to follow the $T_k$ Bland switches of Lemma~\ref{lem:acyclic-counter} in exact order. It remains to verify that Dantzig then terminates and that the game satisfies the stated structural and encoding bounds.

At each level $i\le k$, the two immediate rewards differ by $2^i$ while both undiscounted successor values are integer multiples of $2^{i+1}$. The two action values therefore differ by an odd multiple of $2^i$, so at the optimal source policy every unselected action has undiscounted advantage at most $-1$. With $D=1$, bounds~\eqref{eq:deltabound} and~\eqref{eq:acyclic-sharp-bounds} keep the discount perturbation below $1/8$, leaving every unselected forward advantage negative. By~\eqref{eq:clip}, every unselected target action also has negative advantage, so Dantzig stops after $T_k$ switches.

At the sink gate $g_s$, both branches go to $s$, and the return branch pays strictly more than the zero-reward forward branch. Deleting that dominated return branch preserves all Max strategy values and advantages and reduces the sink SCC to the forced two-cycle $s\leftrightarrow g_s$. For any $u\ne s$, each action from $u$ enters a gate whose return branch comes back to $u$, whereas its forward branch advances to the next level or to $s$ and can never revisit $u$. Hence the SCC containing $u$ is
\begin{equation}\label{eq:small-component}
    C_u=\{u\}\cup\{g_a:\src(a)=u\},
\end{equation}
which contains at most three states because $|\A(u)|\le2$ (Figure~\ref{fig:small-scc}). Each gate has at most two actions and a single internal action, placing the game in $\mathcal C_1$. Control alternates on every edge, every simple cycle pairs one original state with one gate, and deleting the single sink return gives $N=n+m$ and $M=3m-1$, matching~\eqref{eq:small-scc-bound}.

Finally, because~\eqref{eq:acyclic-sharp-bounds} holds for every source policy, Lemmas~\ref{lem:preserve} and~\ref{lem:unique} give every Max strategy a unique all-state stationary Min best response, which goes forward at every selected gate and reaches the sink component within $2H+1$ transitions. By Section~\ref{sec:gate-parameters} and~\eqref{eq:acyclic-sharp-bounds}, the source rewards, gate parameters, and $p$ are nonnegative $O(k)$-bit integers. Setting $\gamma=1-1/p$ yields nonnegative target rewards of $O(k)$ bits each, for an $O(k^2)$-bit total encoding across $O(k)$ actions.
\end{proof}

\subsection{The effect of response reoptimization on value gains}\label{sec:response-mechanism}

After each transferred switch, keeping Min's old response would trap play on the two-cycle through the newly selected gate and amplify the gain by $(1-\gamma^2)^{-1}$, whereas Min's updated best response switches to the forward branch and realizes only the source action's discounted advantage.

For any Max switch $\sigma\to\sigma'$ and an all-state best response $\tau$ to $\sigma$, let $\Delta=W^{\sigma'}-W^\sigma$ denote the actual value gain after Min reoptimizes, and let $\widetilde\Delta=V^{\sigma',\tau}-W^\sigma$ denote the hypothetical gain with $\tau$ held fixed.

\begin{proposition}[Gain reduction under response reoptimization]\label{prop:response-loss}
Suppose Max switches to $a$ at $u$ with $g=G_\sigma(a)>0$, and let $P$ and $\mathbf c$ be the transition matrix and reward vector of $(\sigma',\tau)$. Then the residual $\zeta=\mathbf c+\gamma P W^{\sigma'}-W^{\sigma'}$ is nonnegative and vanishes on $X_{\Max}$. With $\mathbf e_u$ the unit coordinate vector at $u$,
\begin{equation}\label{eq:response-loss}
    \widetilde\Delta=(I-\gamma P)^{-1}g\mathbf e_u,\qquad
    \widetilde\Delta-\Delta=(I-\gamma P)^{-1}\zeta\ge0.
\end{equation}
\end{proposition}
\begin{proof}
Relative to $W^\sigma$, the Bellman residual of $(\sigma',\tau)$ equals $g\mathbf e_u$ because only the action at $u$ changed. Relative to $W^{\sigma'}$, $(\sigma',\tau)$ uses $\sigma'$ on $X_{\Max}$ and attains at least the Bellman minimum on $X_{\Min}$, giving $\zeta\ge0$. Applying $(I-\gamma P)(V^{\sigma',\tau}-v)=\mathbf c+\gamma Pv-v$ to $v=W^\sigma$ and $v=W^{\sigma'}$, together with $(I-\gamma P)^{-1}=\sum_{j\ge0}\gamma^jP^j\ge0$, yields~\eqref{eq:response-loss}.
\end{proof}

\begin{proposition}[Gain at a return gate]\label{prop:response-mechanism}
For a switch to $a:u\to z$ in Theorem~\ref{thm:transfer}, let $\Delta$ and $\widetilde\Delta$ denote the gains of the lifted switch $\widehat\sigma\to\widehat\sigma'$, and put $g=G_{\widehat\sigma}(\widehat a)$. Then $0\le\Delta\le\widetilde\Delta$ componentwise, with
\begin{equation}\label{eq:reoptimized-gain}
    0<\Delta(u)=A_\sigma(a)<\widetilde\Delta(u)=\frac{g}{1-\gamma^2}.
\end{equation}
Under the old response, the combined gain on the two states of the gate cycle is $\widetilde\Delta(u)+\widetilde\Delta(g_a)=g/(1-\gamma)$.
\end{proposition}
\begin{proof}
Lemma~\ref{lem:strict} and~\eqref{eq:response-loss} give $0\le\Delta\le\widetilde\Delta$. Under the old response, Lemma~\ref{lem:unique} keeps play from $u$ on the cycle $u\to g_a\to u$, so~\eqref{eq:response-loss} yields $\widetilde\Delta(u)=g/(1-\gamma^2)=K_a-V_\sigma(u)$ (by~\eqref{eq:clip}) and $\widetilde\Delta(g_a)=\gamma\widetilde\Delta(u)$, giving the stated cycle sum.

Under the new best response, Min goes forward at $g_a$ by Lemma~\ref{lem:preserve}. Because $\sigma'$ is proper, its path from $z$ never returns to $u$ and hence agrees with that of $\sigma$, giving $V_{\sigma'}(z)=V_\sigma(z)$. Value preservation then implies $\Delta(u)=V_{\sigma'}(u)-V_\sigma(u)=A_\sigma(a)$, which is positive by Lemma~\ref{lem:strict} and strictly smaller than $\widetilde\Delta(u)$ since $V_{\sigma'}(u)\le B_V<K_a$.
\end{proof}

In the single-player MDP analysis of \citet{post2015simplex}, discounted circulation along a newly created cycle amplifies the switch gain. In the lifted two-player game, that circulation equals $\widetilde\Delta$, which overstates the realized gain $\Delta$ once Min abandons the return branch (Proposition~\ref{prop:response-mechanism}). At the same time, the presence of the return branch caps candidate advantages before the switch and redirects Dantzig onto Bland's exponential path, as the next comparison highlights.

\begin{proposition}[Dantzig switches with and without return actions]\label{prop:forward-only}
From the initial strategy of Theorem~\ref{thm:small-scc}, Dantzig makes $2^{k+1}-k-2$ switches with the return actions present and exactly $k$ switches when only those return actions are deleted. Moreover, for every fixed Max strategy, the two games have the same value at every original state.
\end{proposition}
\begin{proof}
When all gates must go forward, each source action reduces to a two-step transition with discounted reward $r_a$ and effective discount $\gamma^2$, so values and advantages at original states coincide with those of the $\gamma^2$-discounted source MDP. With return actions present, Lemma~\ref{lem:preserve} gives the exact same values at all original states, while Theorem~\ref{thm:small-scc} gives the exponential switch count.

In the forward-only game, the initial undiscounted values at $\alpha_i$ and $\xi_i$ ($i\le k$) are $2^{k+1}$ and $2^{k+1}-2^i$. Because the two states share the same action set and initially select opposite successors, their switch advantages are $-2^i$ and $2^i$ up to a discount error below $1/8$ by~\eqref{eq:deltabound} and~\eqref{eq:acyclic-sharp-bounds}. Dantzig therefore switches $\xi_k$ to $a_\alpha$ first, leaving both states at level $k$ on their optimal action.

After this step and any subsequent switches of lower-level states $\alpha_i$ to $a_\xi$, write $d_i=V(\xi_i)-V(\alpha_i)$ for the current $\gamma^2$-discounted value difference. Then $d_k=0$, and at any level $i<k$ with $d_{i+1}\ge0$, the action value of $a_\xi$ exceeds that of $a_\alpha$ by $2^i+\gamma^2d_{i+1}>0$. This difference equals $d_i$ if $\alpha_i$ has not yet switched, and $d_i=0$ once both states at level $i$ use $a_\xi$. Backward induction keeps $d_i\ge0$ throughout, so only the $k-1$ states $\alpha_1,\ldots,\alpha_{k-1}$ ever improve, each switching once for $k$ switches in total.
\end{proof}

\subsection{Return gates in a two-player source game}\label{sec:two-player-gates}

The same gating principle applies when the source is itself a two-player game: to prescribe Dantzig's switch order, it suffices to gate only the Max actions that can improve along the target trajectory, because their source advantages already incorporate Min's best response elsewhere. Fix a subset of Max actions to be gated, assign them discount $\gamma^2$ in the source game, and assign discount $\gamma$ to all remaining actions. Replacing each designated action by the return gate of Figure~\ref{fig:gate} produces a target game with uniform discount $\gamma$.

\begin{proposition}[Return gates on a subset of Max actions]\label{prop:game-source-gates}
\textup{(i) Value preservation.} Fix a source Max strategy $\sigma$, let $V_\sigma$ be its exact-response value in the source game, and let $A_\sigma(a)$ be the source advantage of $a$ at its assigned discount. If $K_a\ge V_\sigma(u)$ whenever $\sigma(u)=a$ is gated, then the lifted strategy $\widehat\sigma$ has value $V_\sigma(x)$ at every original state $x$. Every ungated Max action retains its advantage $A_\sigma(a)$, while every gated action $a:u\to w$ has target advantage $\min\{A_\sigma(a),C_\sigma(a)\}$ with $C_\sigma(a)=(1-\gamma^2)(K_a-V_\sigma(u))$.

\textup{(ii) Switch order.} Consider a source trajectory that repeatedly switches the first improving Max action in a fixed priority order, and assume the condition of \textup{(i)} holds at every visited strategy $\sigma$. If at each step every improving Max action $a$ is gated, satisfies $0<C_\sigma(a)\le A_\sigma(a)$, and has caps $C_\sigma(a)$ that strictly decrease in priority order, then Dantzig from $\widehat\sigma_0$ reproduces the trajectory with a unique choice at every switch.
\end{proposition}
\begin{proof}
Evaluated from $u$ at the source values $V_\sigma$, the forward and return branches of a gated action $a:u\to w$ yield $V_\sigma(u)+A_\sigma(a)$ and $V_\sigma(u)+C_\sigma(a)$, as in~\eqref{eq:clip}. When $a=\sigma(u)$, the forward branch equals $V_\sigma(u)$ and the return branch is at least $V_\sigma(u)$. Assigning each gate the minimum of its two branch values divided by $\gamma$ satisfies all Bellman equations of $\widehat\sigma$, including the minima at original Min states, so Lemma~\ref{lem:bellman} gives value preservation and the stated advantages.

Under \textup{(ii)}, every improving Max action attains its positive cap $C_\sigma(a)$, while all other Max actions have nonpositive target advantage by \textup{(i)}. Dantzig therefore uniquely selects the highest-priority improving action at every step.
\end{proof}

Proposition~\ref{prop:game-source-gates} recovers the switch order in the unrestricted-cost discounted counter of \citet[Section~3]{mei2026simplestrategyiterationdantzigsrule}: after negating costs to convert to reward maximization, each level's Max state (denoted $a_i$ there) has return gates on both outgoing actions, whereas its Min state ($b_i$) is ungated. Writing $\beta=\gamma^2$, setting $K_a=K_i/(1-\beta)$ matches Mei's reduced-cost cap $K_i+(1-\beta)v_{a_i}^\pi$ under cost negation, and Mei's value bounds verify \textup{(i)} and \textup{(ii)} so that Dantzig switches the lowest-indexed improving level across all $2^k-1$ steps.

\section{Exact-gain comparison and fixed-depth lookahead}\label{sec:lookahead}

Can Max escape exponential trajectories by evaluating Min's reoptimized best response for every candidate switch and choosing the switch of largest total value gain? A cyclic return-gate gadget rules this out. With two replicas of each source state, switching a single replica leaves Min an escape through the unchanged replica and yields only a tiny gain, whereas switching the second replica unlocks the full source improvement. More generally, with $d+1$ replicas per state, any search of depth $d$ from a strategy in which all copies of each state agree leaves at least one unchanged copy per state. For $d>1$, the last-label tie-breaking rule of Section~\ref{sec:lookahead-rule} ensures that each source switch takes exactly $d+1$ executed switches.

\subsection{Cyclic return gates}\label{sec:cyclic-construction}

For lookahead depth $d\ge1$, set $q=d+1$. Theorem~\ref{thm:common-lower-bound} corresponds to $q=2$.

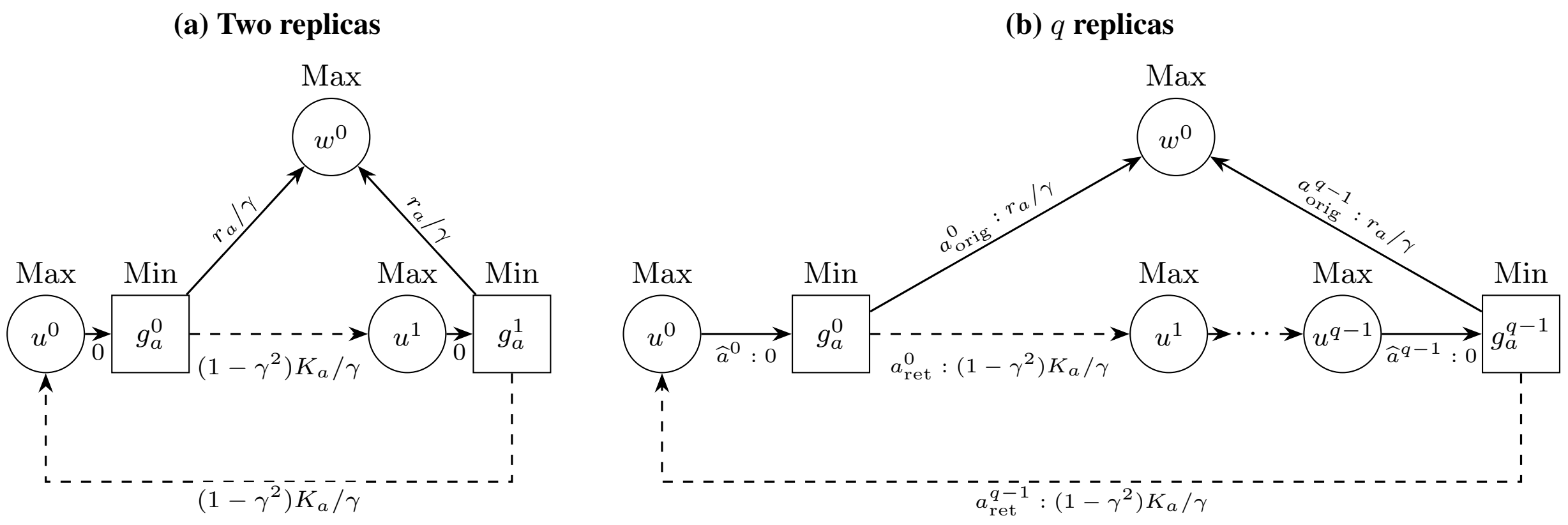
\begin{figure}[htbp]
\centering
\begin{minipage}{\linewidth}
\centering
\vspace*{6pt}
\begin{minipage}[t]{0.35\linewidth}
\centering
\vspace{0pt}
{\small\bfseries (a) Two replicas\endgraf}
\smallskip
\begin{tikzpicture}[x=0.70cm,y=0.85cm,font=\footnotesize,
  state/.append style={minimum size=8mm,text width=6.4mm,align=center,inner sep=0.5pt,font=\fontsize{8}{9}\selectfont},
  gate/.append style={minimum size=8mm,text width=6.4mm,align=center,inner sep=0.5pt,font=\fontsize{8}{9}\selectfont},
  reward/.style={font=\scriptsize,fill=white,inner sep=1pt}]
\node[state,label=above:{$\Max$}] (u0) at (0,0) {$u^0$};
\node[gate,label=above:{$\Min$}] (g0) at (1.55,0) {$g_a^0$};
\node[state,label=above:{$\Max$}] (u1) at (5.35,0) {$u^1$};
\node[gate,label=above:{$\Min$}] (g1) at (6.9,0) {$g_a^1$};
\node[state,label=above:{$\Max$}] (w) at (4.225,2.4) {$w^0$};
\draw[edge] (u0) -- node[reward,below=0.6mm,fill=none] {$0$} (g0);
\draw[edge,dashed] (g0) -- node[reward,below=1.5mm,fill=none] {$(1-\gamma^2) K_a/\gamma$} (u1);
\draw[edge] (u1) -- node[reward,below=0.6mm,fill=none] {$0$} (g1);
\draw[edge,dashed] (g1.south) -- (6.9,-1.8) -- node[reward,below] {$(1-\gamma^2) K_a/\gamma$} (0,-1.8) -- (u0.south);
\draw[edge] (g0) -- node[reward,above,sloped] {$r_a/\gamma$} (w);
\draw[edge] (g1) -- node[reward,above,sloped] {$r_a/\gamma$} (w);
\end{tikzpicture}
\end{minipage}\hfill
\begin{minipage}[t]{0.63\linewidth}
\centering
\vspace{0pt}
{\small\bfseries (b) $q$ replicas\endgraf}
\smallskip
\begin{tikzpicture}[x=1cm,y=0.85cm,font=\footnotesize,
  state/.append style={minimum size=8mm,text width=6.4mm,align=center,inner sep=0.5pt,font=\fontsize{8}{9}\selectfont},
  gate/.append style={minimum size=8mm,text width=6.4mm,align=center,inner sep=0.5pt,font=\fontsize{8}{9}\selectfont},
  reward/.style={font=\fontsize{6.5}{7.5}\selectfont,fill=white,inner sep=1pt}]
\node[state,label=above:{$\Max$}] (u0) at (0,0) {$u^0$};
\node[gate,label=above:{$\Min$}] (g0) at (1.75,0) {$g_a^0$};
\node[state,label=above:{$\Max$}] (u1) at (5.25,0) {$u^1$};
\node at (6.15,0) {$\cdots$};
\node[state,label=above:{$\Max$}] (uq) at (7.05,0) {$u^{q-1}$};
\node[gate,label=above:{$\Min$}] (gq) at (8.90,0) {$g_a^{q-1}$};
\node[state,label=above:{$\Max$}] (w) at (5.325,2.4) {$w^0$};
\draw[edge] (u0) -- node[reward,below=0.6mm,fill=none,inner sep=0.5pt] {$\widehat a^0:0$} (g0);
\draw[edge,dashed] (g0) -- node[reward,below=1.5mm,fill=none] {$a_{\ret}^{0}:(1-\gamma^2) K_a/\gamma$} (u1);
\draw[edge] (u1) -- (5.90,0);
\draw[edge] (6.40,0) -- (uq);
\draw[edge] (uq) -- node[reward,below=0.6mm,fill=none,inner sep=0.5pt] {$\widehat a^{q-1}:0$} (gq);
\draw[edge,dashed] (gq.south) -- (8.90,-1.8) --
  node[reward,below] {$a_{\ret}^{q-1}:(1-\gamma^2) K_a/\gamma$} (0,-1.8) -- (u0.south);
\draw[edge] (g0) -- node[reward,above,sloped] {$a_{\orig}^{0}:r_a/\gamma$} (w);
\draw[edge] (gq) -- node[reward,above,sloped] {$a_{\orig}^{q-1}:r_a/\gamma$} (w);
\end{tikzpicture}
\end{minipage}
\end{minipage}
\caption{Cyclic gates for $a:u\to w$. Forward branches lead to $w^0$, and dashed return branches lead to the next replica, with $u^q=u^0$. Panel (a) shows immediate rewards for two replicas. Panel (b) also names the actions for $q=d+1$ replicas. If $u=w$, then $w^0=u^0$.}\label{fig:cyclic-gate}
\end{figure}

Given a source MDP with $n$ states, $m$ actions, and at most two actions per state, replace each source state $u$ by $q$ Max replicas $u^0,\ldots,u^{q-1}$, with superscripts taken modulo $q$. For each source action $a:u\to w$, replica $u^i$ has a zero-reward action $\widehat a^i$ entering a Min gate $g_a^i$. With target discount $\gamma\in(0,1)$ and gate parameter $K_a>0$, Min's two actions at $g_a^i$ are
\begin{align}\label{eq:cyclic-gates}
    a_{\orig}^{i} &:g_a^i\longrightarrow w^0
       &&\text{with reward }r_a/\gamma,\notag\\
    a_{\ret}^{i} &:g_a^i\longrightarrow u^{i+1}
       &&\text{with reward }(1-\gamma^2)K_a/\gamma.
\end{align}
The return branches thus link the $q$ replicas of $u$ into a directed cycle (Figure~\ref{fig:cyclic-gate}). Control alternates on every transition, each state has at most two actions, and before deleting the dominated sink returns the game has $N=q(n+m)$ states and $M=3qm$ actions.

For a source policy $\sigma$, the \emph{matched strategy} $\widehat\sigma$ selects $\widehat{\sigma(u)}^i$ at every replica $u^i$. Unlike Section~\ref{sec:gate-parameters}, where $K_a$ varied linearly with $\rank(a)$, Section~\ref{app:cyclic-copies} sets $K_a=K_0\Lambda^{m-\rank(a)}$ with geometric ratio $\Lambda=8N^2$ and chooses $\gamma=1-1/p$ close enough to one so that each matched strategy $\widehat\sigma$ on the acyclic counter $\mathcal M_k$ satisfies three properties (verified in~\eqref{eq:cyclic-parameters}--\eqref{eq:cyclic-current-gain} and Lemma~\ref{lem:cyclic-one-copy-gain}):
\begin{enumerate}[label=(\alph*),nosep,leftmargin=*]
\item Every replica preserves the $\gamma^2$-discounted source value, $W^{\widehat\sigma}(u^i)=V_\sigma(u)$, and switching $u^i$ to $\widehat a^i$ has current advantage $G_{\widehat\sigma}(\widehat a^i)=\min\{A_\sigma(a),C_\sigma(a)\}$ with $C_\sigma(a)=(1-\gamma^2)(K_a-V_\sigma(u))$.
\item If $a:u\to w$ is Bland's choice at $\sigma$, then $G_{\widehat\sigma}(\widehat a^i)=C_\sigma(a)>0$ for every $i$, while geometric rank separation ensures $G_{\widehat\sigma}(\widehat b^j)<C_\sigma(a)/N$ for every source action $b\ne a$ and every replica $j$.
\item Because $\mathcal M_k$ has integer rewards, any improving source action satisfies $A_\sigma^1(a)\ge1$, and the discount choice ensures $A_\sigma(a)>1/2$ together with $N^2C_\sigma(a)<1/4$.
\end{enumerate}

\begin{proof}[Proof of Theorem~\ref{thm:common-lower-bound}]
Apply the two-replica construction ($q=2$) with unscaled rewards to the acyclic counter $\mathcal M_k$. At a matched strategy $\widehat\sigma$, write $v=W^{\widehat\sigma}$, let $a:u\to w$ be Bland's next action, and put $g=C_\sigma(a)>0$. By properties (b) and (c), both copies of $a$ have current advantage $g$ and every action $b\ne a$ has advantage strictly below $g/N$, with $N^2g<1/4$ and $A_\sigma(a)>1/2$.

Fix an all-state Min best response $\tau$ to $\widehat\sigma$, which goes forward at the gate of every action used by $\widehat\sigma$. Consider any Max strategy under which every source state still has at least one replica using its $\widehat\sigma$-action. Under $\tau$, consecutive return steps cannot cycle between the two replicas, and every forward step outside the sink advances a counter level or enters the sink, so every play reaches the sink component in at most $N$ transitions. Relative to $v$, the one-step Bellman residual $c(b)+\gamma v(f(b))-v(x)$ along $\tau$ vanishes at all gates and unchanged replicas and equals $G_{\widehat\sigma}(b)$ at each changed replica. Summing these discounted residuals along $\tau$ upper-bounds $V^{\Sigma,\tau}-v$ and hence $W^\Sigma-v$ for any such strategy $\Sigma$.

\emph{The first switch.} If a single replica switches from $\widehat\sigma$ to an action of positive current advantage $\eta$, any play under $\tau$ visits that replica at most once, so every state gains at most $\eta$. By Lemma~\ref{lem:strict}, the switched replica gains at least $\eta$ and no state loses value, placing the exact total gain $F(\Sigma)-F(\widehat\sigma)$ in $[\eta,N\eta]$. Because $\eta=g$ for either copy of $a$ and $\eta<g/N$ for every $b\ne a$, both Dantzig and modified simplex must switch one of the two copies of $a$.

\emph{The second switch.} Let $\Sigma$ be the strategy after switching one copy of $a$, and write $\Delta=W^\Sigma-v$. The switched replica gains $g$, while the unchanged replica of $u$ and $w^0$ retain their values $v$ because plays from them under $\tau$ never visit the switched replica. At the unchanged replica of $u$, switching to the other copy of $a$ yields forward increment $A_\sigma(a)>2g$ and return increment $(1+\gamma^2)g$ (returning through the already-switched replica), so completing the pair has current advantage $(1+\gamma^2)g$. For any other candidate switch $b:x\to y$, including reverting the first switch, $0\le\Delta\le g\mathbf1$ implies
\[
    G_\Sigma(b)=G_{\widehat\sigma}(b)+\gamma\Delta(y)-\Delta(x)
    \le (\gamma+1/N)g < (1+\gamma^2)g,
\]
since $1+\gamma^2-\gamma\ge3/4$ and $N\ge24$. Dantzig therefore switches the second copy of $a$.

Switching that second copy reaches the next matched strategy $\widehat\sigma'$. Because the downstream value at $w$ is unchanged, property (a) increases the value at $u^0$ (and $u^1$) by $A_\sigma(a)>1/2$ over $v$, with all other states nondecreasing, so $F(\widehat\sigma')-F(\widehat\sigma)>1/2$. By contrast, any competing second switch leaves at least one unchanged replica at every source state. Along $\tau$, every pre-sink residual is at most $g$ and sink residuals are zero, so summing at most $N$ steps across all $N$ initial states bounds the total gain over $F(\widehat\sigma)$ by $N^2g<1/4$. Modified simplex therefore also completes the pair.

Inductively, every adaptive combination of Dantzig and modified simplex executes exactly two switches per Bland switch of $\mathcal M_k$, under arbitrary tie breaking. Lemma~\ref{lem:acyclic-counter} gives $2^{k+1}-k-2$ Bland switches, and at the final matched strategy every unselected source action has undiscounted advantage at most $-1$, which remains negative after discounting by~\eqref{eq:Aerror} and~\eqref{eq:cyclic-smallness}, terminating both rules. Deleting the two dominated sink return actions yields $N=12k+12$ and $M=24k+16$, with at most six states per SCC and at most one internal action per Min gate. Scaling rewards by a common positive factor normalizes them to $[0,1]$ without altering either rule and yields an $O(k^2\log(k+1))$-bit encoding (Section~\ref{app:cyclic-copies}).
\end{proof}

When $d>1$, distinct first switches can reach the same optimal depth-$d$ endpoint even when that endpoint completes the current source switch. Section~\ref{sec:lookahead-rule} specifies how tie breaking governs the exact trajectory.

\subsection{Lookahead rule and its scope}\label{sec:lookahead-rule}

Depth-$d$ lookahead evaluates all improving switch paths of length at most $d$ by their endpoint total value $F$, executes only the first switch of an optimal path, and then searches anew to depth $d$ from the updated strategy.

Specifically, each trial step changes the action at a single Max state to one of positive current advantage, and every trial strategy is evaluated against an exact all-state Min best response in the infinite-horizon discounted game. For an integer $\ell\ge0$, let $\mathcal L_\ell(\Sigma)$ denote the maximum of $F$ reachable from $\Sigma$ in at most $\ell$ improving switches. Because every improving switch strictly increases $F$ (Lemma~\ref{lem:gain-sign}),
\begin{equation}\label{eq:lookahead-score}
    \mathcal L_\ell(\Sigma)=
    \begin{cases}
        F(\Sigma), & \text{if $\ell=0$ or no action improves},\\
        \displaystyle\max_{a:\,G_\Sigma(a)>0}\mathcal L_{\ell-1}(\Sigma^a), & \text{otherwise}.
    \end{cases}
\end{equation}
The algorithm stops when no action improves and otherwise executes an improving switch $a$ maximizing $\mathcal L_{d-1}(\Sigma^a)$. For $d=1$, this rule coincides with modified simplex.

For $d>1$, the lower bound below uses \emph{last-label tie breaking}. Every Max action carries a label, with all copies $\widehat a^i$ of a source action $a$ sharing label $a$, and all Max actions are given a fixed tie-breaking total order. Among improving first switches that maximize $\mathcal L_{d-1}(\Sigma^a)$, the rule retains those carrying the label of the most recently executed switch whenever such a candidate exists. Otherwise, including at the initial strategy, it retains those maximizing the one-switch exact gain $F(\Sigma^a)-F(\Sigma)$, and breaks any remaining tie by the fixed total order. Only executed outer switches update the stored label.

On the two-bit counter $\mathcal M_2$ of Figure~\ref{fig:acyclic-example} with $q=3$ replicas, depth-two lookahead under last-label tie breaking executes all four Bland switches across all three replicas, taking $4\cdot3=12$ switches. By contrast, suppose endpoint ties are broken solely by a fixed state-replica priority order that ranks states $\xi_2,\alpha_2,\xi_1,\alpha_1$ and replicas $2,1,0$ within each state, highest priority first. Once $\xi_1^2$ and $\xi_1^1$ have switched to the upper row, switching $\xi_1^0$ and $\xi_2^0$ in either order reaches the same best two-step endpoint. Last-label priority finishes $\xi_1^0$ first, whereas the fixed order selects $\xi_2^0$, finishes $\xi_2$, reverts the two switched $\xi_1$ replicas, and then switches all three $\alpha_1$ replicas. Because $\xi_1^0$ never leaves its initial action, this shortcut reaches the same optimal strategy in $10$ switches instead of $12$. Whether some family forces exponentially many depth-two switches under arbitrary tie breaking remains open.

\begin{theorem}[Exponential lower bounds for fixed-depth lookahead]\label{thm:lookahead-exponential}
For every $d,k\ge1$, there is a deterministic two-player zero-sum discounted game in $\mathcal C_1$ with labeled Max actions and an initial Max strategy from which depth-$d$ lookahead (using last-label tie breaking and any fixed total order when $d>1$, and arbitrary tie breaking when $d=1$) executes $T_{k,d}$ switches, where
\begin{align}\label{eq:lookahead-lower-bound}
    N=(d+1)(6k+6),\qquad M=(d+1)(12k+8),\qquad T_{k,d}=(d+1)(2^{k+1}-k-2).
\end{align}
Every state has at most two actions, every SCC has at most $3(d+1)$ states, and control alternates on every transition. Rewards can be normalized to $[0,1]$, with total binary encoding length $O((d+1)k^2\log((d+1)(k+1)))$.
\end{theorem}

From each matched strategy, every optimal depth-$d$ endpoint switches at least one replica to Bland's next action, and once one replica has switched, every optimal endpoint completes all remaining replicas of that state. Lemma~\ref{lem:cyclic-transfer} shows that last-label tie breaking selects a matching first step throughout, yielding $d+1$ executed switches per Bland switch.

Normalizing rewards to $[0,1]$ bounds their magnitudes while allowing binary precision to grow with $k$ and $d$. A naive depth-$d$ search solves $O(M^d)$ deterministic MDPs per outer step, which takes strongly polynomial arithmetic work per step for each fixed $d$, and Theorem~\ref{thm:lookahead-exponential} counts accepted outer switches even if all candidate evaluations are free. When $d\ge|X_{\Max}|$, Lemma~\ref{lem:restricted-optimality} implies that lookahead can already see a globally optimal strategy, although searching that deep need not take polynomial time and the rule still executes one switch per step.

\begin{remark}[Dependence on depth and discount horizon]
For $k\ge2$, $2^{k+1}-k-2\ge2^k$ gives
\begin{align}\label{eq:lookahead-depth-size}
    T_{k,d}\ge(d+1)2^{N/[6(d+1)]-1}.
\end{align}
As $N\to\infty$ with $d=o(N/\log N)$, this switch count is superpolynomial both in $N$ and in the binary input length $O(N^2\log N)$. For fixed $d$, the geometric gate parameters give $\log h=\Theta(k\log k)$ and hence $T_{k,d}=\exp(\Theta(\log h/\log\log h))$, whereas the sharper $\Omega(h/\log h)$ horizon bound in Corollary~\ref{cor:small-scc-progress} uses the single-replica construction of Section~\ref{sec:transfer}.
\end{remark}

\section{Upper bounds from response and component structure}\label{sec:structure}

The lower bounds of Sections~\ref{sec:main-results}--\ref{sec:lookahead} motivate two complementary questions: when can deterministic discounted games be solved in strongly polynomial arithmetic time without global strategy iteration, and what structural conditions guarantee polynomial switch bounds for global single-state rules? Recall that arithmetic complexity counts elementary operations to compute $W^*$ and an optimal strategy pair, whereas $T$ counts outer Max switches. On the games of Theorem~\ref{thm:small-scc}, backward component solving takes $O(N+M)$ operations even though global Dantzig takes $2^{\Omega(N)}$ switches (Corollary~\ref{cor:small-scc-linear}). Table~\ref{tab:structural-bounds} summarizes the algorithmic and switch upper bounds of this section. All switch bounds hold from any initial Max strategy and under arbitrary tie breaking, using exact all-state Min best responses (whose common value vector $W^\sigma$ makes switch scores independent of which optimal response Min selects).

\begin{table}[!htbp]
\centering
\caption{Arithmetic cost of the component solver and switch upper bounds $T$ for global rules. Here $P(N,M)$ bounds Dantzig pivots in deterministic MDPs, $L=O(N^3M^2\log^2(N+1))$ sums the per-component switch bounds for one fixed-boundary phase, and $h=(1-\gamma)^{-1}$.}\label{tab:structural-bounds}
{\setlength{\tabcolsep}{3pt}
\renewcommand{\arraystretch}{1.05}
\begin{tabularx}{\linewidth}{@{}>{\raggedright\arraybackslash}p{0.35\linewidth}>{\raggedright\arraybackslash}p{0.22\linewidth}>{\raggedright\arraybackslash}p{0.25\linewidth}>{\raggedright\arraybackslash}X@{}}
\toprule
Conditions & Method & Upper bound & Result\\
\midrule
$\mathcal C_1$ & Component solver & $O(N^2M)$ arithmetic operations & Thm.~\ref{thm:component-solver}\\
\addlinespace[2pt]
\leavevmode Forced zero-reward sink. Each other Min state has at most one nonsink action, and $e$ Min states have both. & Global Dantzig & $T\le(e+1)P(N,M)$ & Thm.~\ref{thm:fixed-exits}\\
\addlinespace[2pt]
$\mathcal C_1$, condensation height $\rho$ & Global Dantzig & $T\le(1+L/\rho)^\rho-1$ & Thm.~\ref{thm:component-height}\\
\addlinespace[2pt]
\leavevmode Bounded SCC size and bounded number of actions per state & Single-state Max improvement & $T\le N2^{O(\rho)}$ & Thm.~\ref{thm:bounded-components}\\
\addlinespace[2pt]
\leavevmode Each visited Max strategy admits a best response under which all Max decision states are transient & Dantzig or modified simplex & $T=O(MNh\log(N+1))$ & Cor.~\ref{cor:transient-decisions}\\
\bottomrule
\end{tabularx}}
\end{table}

\subsection{The component solver}\label{sec:component-solver}

Because play can never reenter an SCC after leaving it, processing the SCCs in reverse topological order fixes the optimal continuation value of every exit before its source component is solved. Within a mixed component in $\mathcal C_1$, a Min exit can be eliminated permanently whenever an upper bound on continuing inside the component is already strictly smaller than that exit's fixed payoff.

\begin{theorem}[Component solver for $\mathcal C_1$]\label{thm:component-solver}
For every game in $\mathcal C_1$, the component solver computes $W^*$ and an optimal stationary strategy pair in $O(N^2M)$ arithmetic operations, independently of the discount factor and reward magnitudes, with all intermediate rational numbers of polynomial bit length in the input size.
\end{theorem}

When solving a component $C$, redirect every exit $a:x\to y$ to a local zero-reward sink with immediate reward
\begin{equation}\label{eq:fixed-boundary}
    c(a)+\gamma W^*(y),
\end{equation}
and preprocess each state with exits by keeping a single exit of largest payoff at Max states and smallest payoff at Min states. Any single-player component reduces immediately to a deterministic MDP.

In a mixed component $C$, strong connectivity gives every Min state $x\in C\cap X_{\Min}$ at least one internal action, and membership in $\mathcal C_1$ makes this internal action $a_x$ unique. Initialize Min to take its retained exit wherever available and $a_x$ elsewhere, and solve the resulting deterministic Max MDP for its optimal value vector $U$. Because restricting Min can only raise payoffs, $U$ upper-bounds the local game value on $C$, and $U(x)$ equals the retained exit payoff at every Min state $x$ that currently exits. Whenever such a state satisfies
\[
    c(a_x)+\gamma U(f(a_x))<U(x),
\]
continuing via $a_x$ is strictly better than exiting even under the upper bound $U$, and therefore also under the true local game value. Permanently delete the exit at all such states, switch Min to $a_x$ there, and resolve the Max MDP. Once no remaining exit satisfies this strict inequality, $U$ satisfies all local Bellman equations and hence equals $W^*$ on $C$. After all SCCs have been solved, selecting one Bellman-tight original action per state yields a globally optimal stationary strategy pair.

Sections~\ref{app:exit-elimination} and~\ref{app:component-solver-proof} prove correctness and the $O(N^2M)$ bound, while Section~\ref{sec:branching-solver} extends the solver beyond $\mathcal C_1$ by enumerating internal Min choices at a cost parameterized by internal branching.

\textbf{Periodic stopping.}
Unlike the three-state return-gate components of Section~\ref{sec:transfer}, mixed SCCs in $\mathcal C_1$ can be arbitrarily large. In a periodic zero-sum specialization of the sequential stopping games of \citet{solanvieille2003}, Max and Min alternate around a cycle of an arbitrary even number of phases: stopping pays a phase reward and enters an absorbing terminal, whereas continuing advances to the next phase (wrapping from the last phase back to the first). The continuation edges thus form a single large mixed SCC in which every Min state has a unique internal action. Section~\ref{app:component-solver-proof} gives a second illustration arising from repeated ko fights in Go.

\subsection{Global Dantzig with fixed exit values}

Along a global Dantzig trajectory, $W^\sigma$ never decreases (Lemma~\ref{lem:strict}), so once a fixed exit payoff at a Min state is at most the value of continuing, exiting remains optimal at all subsequent steps.

Let $P(N,M)$ denote a nondecreasing upper bound on Dantzig pivots in deterministic MDPs with at most $N$ states, at most $M$ actions, and a common discount, valid from any initial policy and under arbitrary tie breaking. Negating rewards to convert between cost minimization and reward maximization yields $P(N,M)=O(N^2M^2\log^2(N+1))$ by \citet{hansen2014dantzig}, independently of the rewards and discount.

\begin{theorem}[Dantzig bound with fixed exits]\label{thm:fixed-exits}
Suppose $s$ is a forced zero-reward sink and every other Min state has at most one action leading outside $\{s\}$, with $e$ Min states having both a sink action and a nonsink action. Along any global Dantzig trajectory, all-state Min best responses can be chosen so that only those $e$ states ever change action, each at most once (from continuing to exiting). The trajectory therefore splits into at most $e+1$ fixed-response phases and satisfies
\begin{equation}\label{eq:fixed-exits}
    T\le(e+1)P(N,M)
\end{equation}
for every initial Max strategy, every Max tie-breaking rule, and every exact response solver. Pairing Dantzig with a strongly polynomial deterministic-MDP response solver makes the overall algorithm strongly polynomial.
\end{theorem}

To see why, fix a least-reward sink action at each Min state that has one and select it whenever it is Bellman-optimal, including on ties. Within each resulting fixed-response phase, a single Min strategy $\tau$ remains optimal before every Max switch, so the game values and Max advantages before each switch coincide with those in the deterministic Max MDP induced by $\tau$. The switches in that phase thus form a valid Dantzig prefix of length at most $P(N,M)$ (Lemma~\ref{lem:response-phases}), as proved in Section~\ref{app:fixed-exits-proof}.

\subsection{Condensation height and global updates}

When an exit leads to another nontrivial SCC rather than a fixed sink, a downstream Max switch can raise the exit payoff and cause Min to switch back from exiting to continuing, as the four-state example in Section~\ref{app:fixed-exits-proof} illustrates. Splitting the global trajectory at every Max switch strictly downstream of a component $C$ partitions play into the \emph{fixed-boundary phases} of $C$. Within each such phase, all exit payoffs from $C$ remain constant, allowing Theorem~\ref{thm:fixed-exits} to bound local switches inside any mixed component of a game in $\mathcal C_1$ and compounding across chains of length at most $\rho$.

\begin{theorem}[Global Dantzig bound by condensation height]\label{thm:component-height}
For any game in $\mathcal C_1$ of condensation height $\rho$, global Dantzig with exact all-state responses from any initial Max strategy and under arbitrary tie breaking makes $T$ switches, where
\begin{equation}\label{eq:component-height}
    T\le\left(1+\frac{L}{\rho}\right)^\rho-1,\qquad
    L=O(N^3M^2\log^2(N+1)).
\end{equation}
Here $L$ sums the per-component fixed-boundary switch budgets, each depending only on the states and actions of its component (see~\eqref{eq:height-local-budget} in Section~\ref{app:height-proof}). For every fixed $\rho$, computing exact responses with a strongly polynomial deterministic-MDP solver yields a strongly polynomial implementation of global Dantzig.
\end{theorem}

Even when Min has multiple internal actions, successive strictly improving single-state Max switches inside $C$ during a single fixed-boundary phase can never revisit a local Max strategy. Counting local Max strategies yields a second height-based upper bound without requiring membership in $\mathcal C_1$.

\begin{theorem}[Global improvement with bounded components]\label{thm:bounded-components}
For any deterministic discounted game, set
\begin{align}\label{eq:local-strategy-count}
    S_C=\prod_{x\in C\cap X_{\Max}}|\A(x)|,\qquad S_{\max}=\max_C S_C,
\end{align}
with empty products equal to one. Let $d_{\mathrm{out}}$ be the maximum outdegree of the condensation graph (counting distinct successor components) and $r$ be its number of source components. From any initial Max strategy, every exact-response trajectory that updates one Max state to an action of strictly positive advantage at each step satisfies
\begin{align}\label{eq:bounded-components}
    T\le r(S_{\max}-1)\sum_{j=0}^{\rho-1}(d_{\mathrm{out}}S_{\max})^j.
\end{align}
In particular, if every SCC has at most $n_{\max}$ states and every state has at most $d_{\max}$ actions, then $S_{\max}\le d_{\max}^{n_{\max}}$ and $d_{\mathrm{out}}\le n_{\max}d_{\max}$, giving $T\le N2^{O(\rho)}$ for fixed $n_{\max},d_{\max}$. Consequently, for any fixed constants $n_{\max},d_{\max},c_\rho$, global Dantzig admits a strongly polynomial implementation whenever $\rho\le c_\rho\log(N+1)$.
\end{theorem}

Theorem~\ref{thm:bounded-components} applies under arbitrary tie breaking, with $T=0$ when $S_{\max}=1$ and only the $j=0$ term present in~\eqref{eq:bounded-components} when $d_{\mathrm{out}}=0$. Section~\ref{app:height-proof} proves Theorems~\ref{thm:component-height} and~\ref{thm:bounded-components}. Because the family in Theorem~\ref{thm:small-scc} has $n_{\max}=3$, $d_{\max}=2$, $\rho=k+2$, and $T=2^{k+1}-k-2$, exponential dependence on $\rho$ is unavoidable even under bounded local parameters, although the base of the exponent need not be tight.

\subsection{Switch bounds from discounted visits to Max decision states}\label{sec:response-occupation}

Write $J=\{u\in X_{\Max}:|\A(u)|\ge2\}$ for the set of Max decision states. Under any stationary pair, launch one play from each of the $N$ states and weight every visit to $J$ at step $j\ge0$ by $\gamma^j$. If the resulting total discounted visitation of $J$ under an optimal response is bounded by $B$ while the total optimality gap is $E>0$, some currently selected Max action must incur a Bellman optimality loss of at least $E/B$ and can never reappear once the total gap shrinks below $E/B$.

\begin{theorem}[Switch bound from discounted visits]\label{thm:response-occupation}
Consider Dantzig or modified simplex with exact all-state responses from any initial Max strategy and under arbitrary tie breaking. Suppose there exists $B\ge1$ such that every visited Max strategy admits an all-state Min best response under which the discounted number of visits to $J$, summed over the $N$ plays starting from all states, is at most $B$. Then
\begin{equation}\label{eq:response-occupation-bound}
    T\le \left\lceil Nh\log(2B)\right\rceil
    \sum_{u\in J}\bigl(|\A(u)|-1\bigr).
\end{equation}
\end{theorem}

Each switch contracts the total optimality gap by at least a factor of $1-1/(Nh)$. Unless termination occurs earlier, $\lceil Nh\log(2B)\rceil$ switches reduce the gap from $E$ below $E/(2B)$ and permanently eliminate at least one suboptimal Max action, out of at most $\sum_{u\in J}(|\A(u)|-1)$ such actions in total. Section~\ref{app:response-occupation-proof} proves Theorem~\ref{thm:response-occupation} by refining the action-elimination analysis of \citet{jia2020towards}.

\begin{corollary}[Transient Max decision states]\label{cor:transient-decisions}
Suppose every visited Max strategy admits an all-state Min best response such that no directed cycle of the induced play graph intersects $J$. Then both Dantzig and modified simplex satisfy
\begin{equation}\label{eq:transient-decisions-bound}
    T=O\bigl(MNh\log(N+1)\bigr).
\end{equation}
\end{corollary}

When $J$ is transient under a best response, each of the $N$ plays visits each state of $J$ at most once, permitting $B=N|J|$ whenever $J\ne\varnothing$. Without transience, bounding total discounted time across all $N$ plays gives $B=Nh$ and recovers $T=O(MNh\log(2Nh))$, so transience removes $h$ from the logarithmic factor. The lower-bound games in Theorem~\ref{thm:small-scc} satisfy Corollary~\ref{cor:transient-decisions}, yielding an upper bound of $O(h(\log h)^2\log\log h)$ switches to termination at the scale $N,M=\Theta(\log h)$ of Corollary~\ref{cor:small-scc-progress}.

\section{Conclusion}

Global Dantzig strategy iteration can require $2^{\Omega(N)}$ switches on deterministic two-player discounted games that are solvable in $O(N+M)$ arithmetic operations by the backward component solver. When SCC size and action degree are bounded, global Dantzig makes at most $N2^{O(\rho)}$ switches as a function of the condensation height $\rho$, making $\rho=O(\log N)$ sufficient for strong polynomiality, while the lower-bound family with $\rho=\Theta(N)$ shows that exponential dependence on $\rho$ is necessary.

On that same family at discount horizon $h$ and size $N,M=\Theta(\log h)$, even halving the total optimality gap requires $\Omega(h/\log h)$ Dantzig switches. Moreover, on the two-replica family, every run that chooses adaptively at each step between Dantzig's rule and modified simplex takes exponentially many switches under arbitrary tie breaking.

Two concrete questions remain open. First, does depth-two lookahead still require exponentially many switches in the worst case under arbitrary tie breaking? Second, when every visited Max strategy admits a best response under which all Max decision states are transient, what is the sharp dependence of Dantzig's switch count on the discount horizon $h$?

\subsection*{Use of generative AI}
The authors used GPT-6 in OpenAI Codex to polish the prose and correct the grammar. The authors take responsibility for the final text and mathematical conclusions.

\clearpage
\appendix
\renewcommand{\theHsection}{appendix.\Alph{section}}
\renewcommand{\theHsubsection}{\theHsection.\arabic{subsection}}
\renewcommand{\theHsubsubsection}{\theHsubsection.\arabic{subsubsection}}
\pdfbookmark[0]{Appendices}{appendices}

\section{Proof of the exact switch count for the acyclic counter}\label{app:acyclic-counter-proof}

\begin{proof}[Proof of Lemma~\ref{lem:acyclic-counter}]
When bit $j$ is the lowest zero bit of $\nu$, Bland sets it by switching $\xi_j$ and then resets the $j-1$ lower bits by switching $\alpha_{j-1},\ldots,\alpha_1$ in order, making $j$ switches in total.

At every level $i\le k$, both undiscounted successor values in~\eqref{eq:acyclic-source} are multiples of $2^{i+1}$, so the two action values at level $i$ differ by an odd multiple of $2^i$. Backward induction from level $k+1$ therefore determines a unique canonical policy (Section~\ref{sec:acyclic-source}) for each $\nu\in\{0,\ldots,2^k-1\}$. Under the initial policy, $\alpha_i$ and $\xi_i$ have undiscounted values $2^{k+1}$ and $2^{k+1}-2^i$, confirming that it is canonical for $\nu=0$.

At the canonical policy for $\nu<2^k-1$, the states with a positive-advantage action are precisely the states $\xi_i$ with $\nu_i=0$. Bland selects the lowest-ranked among them, $\xi_j$, increasing $V_\sigma^1(\xi_j)$ by at least $2^j$ to match $V_\sigma^1(\alpha_j)$. If $j>1$, equal continuation values at level $j$ make both states at level $j-1$ prefer $a_\xi$ to their current choice $a_\alpha$ by $2^{j-1}$. Inductively, at each affected level $i$ from $j-1$ down to $1$, the rank ordering selects $\alpha_i$ ahead of $\xi_i$ and ahead of any pending higher-level switch, raising $V^1(\alpha_i)$ by $2^i$ above $V^1(\xi_i)$. Whenever $i>1$, both states at level $i-1$ still use $a_\xi$ (since their level-$i$ successors previously had equal value) and now gain $2^i-2^{i-1}=2^{i-1}$ by switching to $a_\alpha$, continuing the cascade down to level $1$.

The resulting policy is canonical for $\nu+1$, and at $\nu=2^k-1$ no improving action remains. Because the $k-j$ bits above $j$ are arbitrary, there are $2^{k-j}$ increments of $j$ switches each, establishing~\eqref{eq:acyclic-count}.
\end{proof}

\section{Proof of the gap and horizon bounds}\label{app:progress-proof}

\begin{proof}[Proof of Corollary~\ref{cor:small-scc-progress}]
In the source counter $\mathcal M_k$, both actions at $\xi_k$ lead to level-$(k+1)$ states with forced moves, so the gap at $\xi_k$ remains constant until the first carry into bit $k$ switches $\xi_k$ at step $t_*$. Put $R=2^{k+1}$. The bounds~\eqref{eq:acyclic-sharp-bounds} give $Q=4(k+1)(6R+1)$, so any horizon $h\ge\bar p_k=16Q$ satisfies Theorem~\ref{thm:transfer}. By Section~\ref{app:acyclic-counter-proof}, $\xi_k$ first switches during the increment from $2^{k-1}-1$ to $2^{k-1}$, which starts at switch $t_*=2^k-k$. At source discount $\gamma^2>1/2$, the two action values at $\xi_k$ are $R/2$ and $\gamma^2R$. Because Lemma~\ref{lem:preserve} preserves every policy value (and hence the optimal value) at all original states, $e_t(\xi_k)=R(\gamma^2-1/2)>0$ for $t<t_*$ and $e_{t_*}(\xi_k)=0$. Monotonicity keeps $e_t(\xi_k)=0$ thereafter, proving~\eqref{eq:small-scc-half} for every $0<\theta<1$.

For the total gap, monotonicity reduces~\eqref{eq:small-scc-total-half} to verifying $E_{t_*-1}>E_0/2$. This is immediate for $k=1$, where $t_*=1$ and $E_0>0$. For $k\ge2$, compare the initial policy $\sigma_0$ with the pre-carry policy $\sigma^-=\sigma_{t_*-1}$ (canonical for $2^{k-1}-1$) via undiscounted reference gaps. At each original state $u$, define $\bar e_\sigma(u)=V^{*,1}(u)-V_\sigma^1(u)$, where $V^{*,1}$ is the optimal undiscounted source value, and at the gate of $a:u\to w$ define
\[
\begin{aligned}
    \bar e_\sigma(g_a)
    &=r_a+V^{*,1}(w)-\min\{r_a+V_\sigma^1(w),V_\sigma^1(u)\}\\
    &=\bar e_\sigma(w)+\max\{0,A_\sigma^1(a)\}.
\end{aligned}
\]
Write $\bar E(\sigma)=\sum_{x\in X}\bar e_\sigma(x)$.

Backward evaluation gives $V^{*,1}(\alpha_i)=V^{*,1}(\xi_i)=3R/2-2^i$ for all $i\le k$. Under $\sigma_0$, the values at $\alpha_i$ and $\xi_i$ are $R$ and $R-2^i$, whereas under $\sigma^-$ both equal $5R/4-2^i$ for $i<k$ and take values $R$ and $R/2$ at level $k$. Hence the two original states at level $i$ contribute $R-2^i$ to $\bar E(\sigma_0)$ and $R/2$ to $\bar E(\sigma^-)$.

Every state at level $i+1$ is the forward successor of two gates from level $i$, so the four gate gaps at level $i$ sum to twice the level-$(i+1)$ original-state gap sum plus the positive undiscounted advantages at level $i$. Under $\sigma_0$, each level $i\le k$ has a single positive advantage $2^i$ (at $\xi_i$), whereas under $\sigma^-$ the only positive advantage is $R/2$ at $\xi_k$. Consequently, the four gates at level $i<k$ contribute $2R-3\cdot2^i$ under $\sigma_0$ and $R$ under $\sigma^-$, those at level $k$ contribute $R/2$ under both policies, and all level-$(k+1)$ and sink gaps vanish. Summing over all states gives
\begin{equation}\label{eq:total-gap-reference}
    \bar E(\sigma_0)=(3k-4)R+8,\qquad
    \bar E(\sigma^-)=(3k-1)R/2,
\end{equation}
so $\bar E(\sigma^-)-\bar E(\sigma_0)/2=3R/2-4$.

To bound the discount perturbation, put $\delta=1-\gamma^2$. Section~\ref{sec:acyclic-source} gives $0\le V_\sigma^1-V_\sigma\le 2HR\delta\,\mathbf1$, which extends to $V^{*,1}-V^*$ by maximizing over policies. Thus each original-state gap differs from $\bar e_\sigma(u)$ by at most $2HR\delta$, and replacing $V_\sigma^1$ with $\gamma^2V_\sigma$ changes any continuation value by at most $(2HR+B_V)\delta\le B_A\delta$. In~\eqref{eq:Wgate} for $\gamma W^{\widehat\sigma}(g_a)$, the forward branch differs from $r_a+V_\sigma^1(w)$ by at most $B_A\delta$, and the return branch $\gamma^2V_\sigma(u)+\delta K_a$ differs from $V_\sigma^1(u)$ by at most $\delta K_a$ (since $K_a>B_A$), so their minimum differs by at most $\delta K_{\max}$. At the source optimum, all advantages are nonpositive and all return caps are positive, so $\gamma W^*(g_a)$ takes the forward branch and differs from its reference by at most $B_A\delta$.

Because $0\le\bar e_\sigma(g_a)\le B_V$, dividing by $\gamma$ and applying~\eqref{eq:Q} and~\eqref{eq:deltabound} gives, for $\sigma=\sigma_t$,
\[
    |e_t(g_a)-\bar e_\sigma(g_a)|
    \le \frac{\delta(K_{\max}+B_A)+(1-\gamma)B_V}{\gamma}
    < \frac{\delta Q}{\gamma}<\frac14.
\]
The original-state gap error is likewise below $1/4$, yielding $|E_t-\bar E(\sigma_t)|<N/4$. Combining the perturbations of $E_{t_*-1}$ and $E_0/2$ adds at most $N/4+N/8=3N/8$. With $N=6k+6$ and~\eqref{eq:total-gap-reference},
\[
    E_{t_*-1}-\frac{E_0}{2}
    >\frac{3R}{2}-4-\frac{3N}{8}
    =3\cdot2^k-4-\frac94(k+1)>0
\]
for all $k\ge2$, proving $T^{\mathrm{sum}}_{1/2}\ge t_*$.

Finally, for any rational $h\ge\bar p_1$, take the largest integer $k$ with $\bar p_k\le h$. Because $\bar p_k=\Theta(k2^k)$ and $\bar p_{k+1}<3\bar p_k$, this choice gives $N,M=\Theta(\log h)$ and $t_*=\Theta(h/\log h)$, establishing~\eqref{eq:small-scc-horizon-bound}. Since every best-response play reaches the forced sink component, Corollary~\ref{cor:transient-decisions} also bounds total switches to termination by $O(h(\log h)^2\log\log h)$ at this scale.
\end{proof}

\section{Cyclic replication for exact-gain and lookahead lower bounds}\label{app:cyclic-copies}

With $q=d+1$ replicas per state, any depth-$d$ search from a matched strategy leaves at least one unchanged replica at every source state, whereas once one replica has switched, the next search can switch the remaining $d$ replicas of that state. Geometric rank spacing gives Bland's next action the largest one-switch exact gain and keeps every incomplete update strictly below the gain of completing that source switch.

\subsection{Improving paths and optimal lookahead endpoints}

Although lookahead restricts attention to sequences of strictly improving switches, a depth-$d$ search still attains the maximum of $F$ over the entire Hamming ball of radius $d$, where $d_H(\Sigma,\Theta)=|\{x\in X_{\Max}:\Sigma(x)\ne\Theta(x)\}|$ denotes Hamming distance~\citep{hamming1950error}.

\begin{lemma}[Optimal values attainable by improving paths]\label{lem:restricted-optimality}
Let $\Sigma^*$ be an all-state optimal Max strategy. From any Max strategy $\Sigma$, at most $d_H(\Sigma,\Sigma^*)$ improving switches suffice to reach a strategy attaining $W^{\Sigma^*}$, each adopting $\Sigma^*$'s action at a state where the current strategy still differs from $\Sigma^*$. Consequently, $\mathcal L_d(\Sigma)=\max_{d_H(\Sigma,\Theta)\le d}F(\Theta)$, and every maximizer $\Theta$ is all-state optimal in the restricted game allowing Max only $\{\Sigma(x),\Theta(x)\}$ at each $x\in X_{\Max}$.
\end{lemma}
\begin{proof}
If $v=W^\Sigma\ne W^{\Sigma^*}$ and no action of $\Sigma^*$ had positive advantage at $v$, then $\mathcal T_{\Sigma^*}v\le v$, and iterating the monotone contraction $\mathcal T_{\Sigma^*}$ would give $W^{\Sigma^*}\le v$, a contradiction. Thus some action of $\Sigma^*$ strictly improves $v$ (Lemma~\ref{lem:strict}) while reducing the Hamming distance to $\Sigma^*$ by one.

Given any $\Theta$ with $d_H(\Sigma,\Theta)\le d$, restrict Max at each state to $\{\Sigma(x),\Theta(x)\}$. An all-state optimal strategy of this restricted game differs from $\Sigma$ on at most $d$ states, dominates $W^\Theta$ componentwise, and is reachable from $\Sigma$ in at most $d$ improving switches by the first part. Conversely, every improving path of length at most $d$ ends inside the Hamming ball. If $\Theta$ maximizes $F$ over the ball, componentwise dominance forces its value vector to equal the restricted optimal value at every state.
\end{proof}

\subsection{Return gates with cyclic replication}

So long as every source state retains at least one unchanged replica, Min can follow return branches at switched replicas until reaching an unchanged replica and thereby cap the total gain. Consider a deterministic total-reward source MDP with $n$ states, $m$ ranked actions, at most two actions per state, and a sink $s$ with a forced zero-reward self-loop. For integers $D,R\ge1$, assume $r_a\in D^{-1}\Z$ and $|r_a|\le R$ for all $a$, and assume every nonterminal simple cycle has total reward at most $-1/D$. Let $\sigma_0,\ldots,\sigma_T$ be a Bland prefix starting from a proper policy $\sigma_0$. Because any cycle created by a single switch from a proper policy has total reward equal to that switch's undiscounted advantage, every switch of nonnegative undiscounted advantage preserves properness, so every policy on the prefix is proper.

Form the $q$-replica game and matched strategies $\widehat\sigma$ of Section~\ref{sec:cyclic-construction}, with $q=d+1$ and $N=q(n+m)$. Call a Max strategy $\Sigma$ \emph{incomplete} relative to $\sigma$ if every source state retains at least one replica using its action in $\widehat\sigma$. For an integer $H\ge n$ bounding absorption steps along the prefix, set $B_V=HR$ and $B_A=H^2R$, and choose
\begin{align}\label{eq:cyclic-parameters}
\begin{alignedat}{3}
    \Lambda&=8N^2, &\qquad K_0&=8N^2(B_V+B_A+1), &\qquad K_a&=K_0\Lambda^{m-\rank(a)},\\
    K_{\max}&=\max_a K_a, &\qquad L_*&=K_{\max}+B_V+B_A+1, &\qquad p&=1024DN^2(q+H+1)L_*,\\
    \gamma&=1-1/p.
\end{alignedat}
\end{align}

Put $\delta=1-\gamma^2$. For every proper policy $\sigma$, Lemma~\ref{lem:perturb} and the lattice gap $1/D$ imply
\begin{align}\label{eq:cyclic-smallness}
    \gamma^N&>1/2,\qquad N^2\delta L_*<\frac{1}{512D(q+H+1)},\notag\\
    A_\sigma(a)&\ge\frac1D-\delta B_A>q\delta (K_{\max}+B_V),
\end{align}
where the second line holds whenever $A_\sigma^1(a)>0$.

Write $V=V_\sigma$ for the $\gamma^2$-discounted source value. At each gate selected by $\widehat\sigma$, the forward branch discounts to $V(u)$ whereas the return branch discounts to $V(u)+\delta(K_a-V(u))>V(u)$, so Bellman uniqueness gives, for $v=W^{\widehat\sigma}$ and every $a:u\to w$,
\begin{align}\label{eq:cyclic-preservation}
    v(u^i)&=V(u),\notag\\
    v(g_a^i)&=\gamma^{-1}\min\{r_a+\gamma^2 V(w),\delta K_a+\gamma^2 V(u)\}.
\end{align}
Consequently, switching any replica $u^i$ to $\widehat a^i$ has current advantage
\begin{align}\label{eq:cyclic-current-gain}
    g_\sigma(a):=G_{\widehat\sigma}(\widehat a^i)=\min\{A_\sigma(a),C_\sigma(a)\},\qquad
    C_\sigma(a)=\delta (K_a-V(u)).
\end{align}

\begin{lemma}[Value bounds for incomplete updates]\label{lem:incomplete-bound}
For $\Sigma$ incomplete relative to a proper policy $\sigma$, let $\mathcal P$ be the set of source actions $a$ with $A_\sigma^1(a)>0$ used at switched replicas of $\Sigma$, and set
\[
    L_\Sigma=B_V+B_A+1+\max\bigl(\{K_a:a\in\mathcal P\}\cup\{0\}\bigr).
\]
Then
\begin{align}\label{eq:incomplete-bound}
    W^\Sigma\le v+N\delta L_\Sigma\mathbf1,\qquad
    F(\Sigma)\le F(\widehat\sigma)+N^2\delta L_\Sigma.
\end{align}
\end{lemma}
\begin{proof}
Pair $\Sigma$ with an all-state Min best response to $\widehat\sigma$ to form a joint stationary policy $\pi$. By~\eqref{eq:cyclic-smallness} and~\eqref{eq:cyclic-current-gain}, Min returns at gates with $A_\sigma^1(a)>0$ and goes forward at all others, since $A_\sigma^1(a)\le0$ implies $A_\sigma(a)\le\delta B_A<C_\sigma(a)$. The one-step residual $\eta=c_\pi+\gamma P_\pi v-v$ vanishes at all gates and unchanged replicas, and at a changed replica equals $C_\sigma(a)$ if Min returns and $A_\sigma(a)$ otherwise. Every positive coordinate of $\eta$ is therefore at most $\delta L_\Sigma$, while any changed replica with $A_\sigma^1(a)<0$ has $\eta\le-1/D+\delta B_A\le-1/(2D)$.

Because every source state retains an unchanged replica whose selected gate goes forward, returns alone cannot form a cycle. Contracting consecutive return steps projects any nonterminal cycle of $\pi$ onto a closed walk in the source MDP, whose total reward is at most $-1/D$ by simple-cycle decomposition. Since source values telescope around a closed walk, the undiscounted advantages along its edges also sum to at most $-1/D$, forcing at least one forward edge to leave a changed replica with $A_\sigma^1(a)<0$. Starting from any state on the cycle in $\pi$, one full traversal therefore accumulates a discounted residual of at most $-\gamma^N/(2D)+N\delta L_\Sigma<0$ by~\eqref{eq:cyclic-smallness}, while the sink cycle has zero residual. Only the transient prefix before entering a cycle (at most $N$ steps) can contribute positively, giving $W^\Sigma\le V^\pi\le v+N\delta L_\Sigma\mathbf1$ and summing to the stated bound on $F(\Sigma)$.
\end{proof}

\subsection{Exact gains of single-replica changes}

When only one replica switches to an improving action, Min can prevent play from visiting that replica more than once, so its current advantage bounds the gain at every state and a factor of $N$ separates competing labels.

\begin{lemma}[Exact gain of a single-replica switch]\label{lem:cyclic-one-copy-gain}
Let $a$ be Bland's choice at $\sigma$. For any source action $b:z\to y$, let $\Sigma_b^i$ change only $z^i$ in $\widehat\sigma$ to $\widehat b^i$. If $g_\sigma(b)>0$, then
\begin{align}\label{eq:cyclic-one-copy}
0\le W^{\Sigma_b^i}-v\le g_\sigma(b)\mathbf1,\qquad
 g_\sigma(b)\le F(\Sigma_b^i)-F(\widehat\sigma)\le N\,g_\sigma(b).
\end{align}
Moreover, $g_\sigma(a)>N g_\sigma(b)$ for every $b\ne a$, so any replica of $a$ achieves a strictly larger one-switch exact gain than every action carrying a different label.
\end{lemma}
\begin{proof}
By Lemma~\ref{lem:perturb} and~\eqref{eq:cyclic-smallness}, $g_\sigma(b)>0$ implies $A_\sigma^1(b)\ge0$, so both $\sigma$ and the switched source policy are proper. Consequently, the $\sigma$-paths from $f(\sigma(z))$ and from $y$ never revisit $z$. Following those forward paths lets Min hold $z^{i+1}$ and $y^0$ down to their $\widehat\sigma$-values, while Lemma~\ref{lem:strict} gives the reverse inequality, leaving both values unchanged and yielding
\begin{align}\label{eq:cyclic-one-replica-values}
    W^{\Sigma_b^i}(z^i)-V(z)=\min\{A_\sigma(b),C_\sigma(b)\}=g_\sigma(b).
\end{align}
Under an optimal Min response to $\widehat\sigma$, both branches at $g_b^i$ enter continuations that never revisit $z^i$, making $g_\sigma(b)$ at $z^i$ the sole nonzero residual relative to $v$. Every state therefore gains at most $g_\sigma(b)$, proving~\eqref{eq:cyclic-one-copy}.

For rank separation,~\eqref{eq:cyclic-parameters} gives
\begin{align}\label{eq:cyclic-rank-separation}
    K_0-B_V>N^2(B_V+B_A+1),\qquad K_a-B_V>N^2(K_b+B_V+B_A+1)\quad\text{whenever }\rank(a)<\rank(b).
\end{align}
By~\eqref{eq:cyclic-smallness}, $g_\sigma(a)=C_\sigma(a)\ge\delta(K_a-B_V)$. Any $b\ne a$ with $A_\sigma^1(b)>0$ has $\rank(b)>\rank(a)$ and $g_\sigma(b)=C_\sigma(b)\le\delta(K_b+B_V)$, whereas $A_\sigma^1(b)=0$ gives $g_\sigma(b)\le\delta B_A$ and $A_\sigma^1(b)<0$ gives $g_\sigma(b)<0$. In all three cases,~\eqref{eq:cyclic-rank-separation} implies $g_\sigma(a)>N g_\sigma(b)$, and~\eqref{eq:cyclic-one-copy} together with Lemma~\ref{lem:gain-sign} yields the strict ordering of exact gains.
\end{proof}

\subsection{Simulation of Bland trajectories under bounded lookahead}

Recall from Section~\ref{sec:lookahead-rule} that all replicas $\widehat a^i$ carry label $a$. At each matched strategy, one-switch exact gain selects Bland's next label, and once its first replica switches, last-label tie breaking completes the remaining $q-1$ replicas one by one.

\begin{lemma}[Bland simulation under bounded lookahead]\label{lem:cyclic-transfer}
Starting from $\widehat\sigma_0$, depth-$d$ lookahead (under last-label tie breaking for $d>1$ and arbitrary tie breaking for $d=1$) reaches $\widehat\sigma_j$ after exactly $qj$ executed switches for every $0\le j\le T$, switching all $q$ replicas of the affected state to Bland's chosen action during each block of $q$ steps.
\end{lemma}
\begin{proof}
\emph{The first replica.} Let $a:u\to w$ be Bland's choice at $\sigma$. Every strategy within $d=q-1$ switches of $\widehat\sigma$ is incomplete relative to $\sigma$. If such an endpoint uses no replica of $a$, every action in $\mathcal P$ has rank larger than $\rank(a)$, so Lemma~\ref{lem:incomplete-bound} and~\eqref{eq:cyclic-rank-separation} bound its total gain over $F(\widehat\sigma)$ strictly below $\delta(K_a-B_V)$, which a single switch to $a$ already achieves. Every optimal depth-$d$ endpoint $\Theta$ therefore uses $a$ at some replica of $u$.

Switching that replica first leaves at most $d-1$ disagreements with $\Theta$ in the game restricted to $\{\widehat\sigma(x),\Theta(x)\}$, so Lemma~\ref{lem:restricted-optimality} supplies an improving path of length at most $d-1$ attaining $W^\Theta$. Label $a$ thus attains the optimal endpoint score $\mathcal L_d(\widehat\sigma)$. Because any previously executed label is already used at all $q$ of its replicas in $\widehat\sigma$, no candidate carries the last label, and Lemma~\ref{lem:cyclic-one-copy-gain} selects a replica of $a$ via one-switch exact gain.

\emph{Completing the switch.} First verify that any proper subset $I$ of replicas of $u$ can switch to $a$ via strictly improving steps. At the resulting strategy $\Sigma$, values dominate $v$ componentwise, while Min can still hold $w^0$ and every unchanged replica of $u$ at their $\widehat\sigma$-values because the $\sigma$-paths from both $f(\sigma(u))$ and $w$ avoid $u$. At any unchanged replica of $u$, switching to $a$ therefore offers forward increment $A_\sigma(a)>0$ and return increment at least $C_\sigma(a)>0$, proving the claim by induction from $I=\varnothing$. Moreover, along a maximal run of $\ell_i$ already-switched consecutive replicas starting at $u^i$ ($1\le\ell_i\le q-1$), returns exit at an unchanged replica and gain at most $\ell_i C_\sigma(a)<q C_\sigma(a)<A_\sigma(a)$ by~\eqref{eq:cyclic-smallness}, so backward induction along the run gives
\begin{align}\label{eq:cyclic-partial}
    W^\Sigma(u^i)-V(u)
       =C_\sigma(a)\sum_{j=0}^{\ell_i-1}\gamma^{2j},
       \qquad 1\le \ell_i\le q-1.
\end{align}

Once $r=|I|$ replicas have switched ($1\le r<q$), switching the remaining $q-r\le d$ replicas reaches the next matched strategy $\widehat\sigma'$. Since the $\sigma$-path from $w$ avoids $u$, it follows that $V_{\sigma'}(u)-V(u)=A_\sigma(a)$, and monotonicity together with~\eqref{eq:cyclic-smallness} gives
\begin{align}\label{eq:cyclic-completion}
    F(\widehat\sigma')-F(\widehat\sigma)
       \ge A_\sigma(a)>1/(2D).
\end{align}
Any depth-$d$ endpoint from $\Sigma$ that leaves $u$ incomplete also cannot complete all $q$ replicas of another source state within $d<q$ switches, so it remains incomplete relative to $\sigma$ and gains at most $N^2\delta L_*<1/(4D)$ over $F(\widehat\sigma)$ by Lemma~\ref{lem:incomplete-bound}. Every optimal endpoint $\Theta$ from $\Sigma$ must therefore complete all $q$ replicas of $a$. Switching one of the remaining $q-r$ replicas of $a$ agrees with $\Theta$ and leaves at most $d-1$ further disagreements (as $\Theta$ agrees with $\Sigma$ on the $r$ already-switched replicas), so Lemma~\ref{lem:restricted-optimality} again yields an improving continuation to $W^\Theta$. Last-label priority therefore executes another copy of $a$ when $d>1$, while for $d=1$ the sole remaining copy of $a$ uniquely achieves the best endpoint. Inducting over $r=1,\ldots,q-1$ and across the $T$ source switches completes the proof.
\end{proof}

\subsection{Proof of Theorem~\ref{thm:lookahead-exponential}}

\begin{proof}
Instantiating the construction on $\mathcal M_k$ gives $n=2k+3$, $m=4k+3$, $H=n$, $R=2^{k+1}$, $D=1$, and $T_k=2^{k+1}-k-2$ Bland switches (Lemma~\ref{lem:acyclic-counter}), with acyclicity vacuously satisfying the negative-cycle condition. Lemma~\ref{lem:cyclic-transfer} yields $q=d+1$ executed switches per Bland switch, and at the final policy every unselected source action has undiscounted advantage at most $-1$, which stays negative after discounting by~\eqref{eq:Aerror} and~\eqref{eq:cyclic-smallness}. By~\eqref{eq:cyclic-current-gain}, no improving switch remains at $\widehat\sigma_{T_k}$, proving $T_{k,d}=qT_k$.

At each of the $q$ sink gates, the zero-reward forward branch is uniquely optimal under every Max strategy, so deleting the $q$ sink return actions preserves all values and lookahead scores while giving $N=q(n+m)$ and $M=q(3m-1)$. For each $u\ne s$, the cyclic returns bind the $q$ replicas of $u$ and their outgoing gates into a single SCC of at most $3q$ states, and acyclicity of $\mathcal M_k$ makes every forward branch an exit, leaving at most one internal action per Min gate and placing the game in $\mathcal C_1$. At the sink, $s^0\leftrightarrow g_s^0$ is the only nonsingleton SCC. Control alternates on every transition, and every state has at most two actions.

Finally,~\eqref{eq:cyclic-parameters} gives $\log_2 K_{\max}=O(m\log N+\log(B_V+B_A+1))=O(k\log(q(k+1)))$, and the same bit-length bound applies to $p$ and every action reward's numerator and denominator, giving an $O(qk^2\log(q(k+1)))$-bit total encoding across $O(qk)$ actions. Normalizing the nonnegative rewards to $[0,1]$ scales all lookahead scores and one-switch gains by the same positive factor, preserving their rankings, ties, and hence the switch trajectories.
\end{proof}

\section{An application to a cyclic Bland counter}\label{sec:source}

Even in the cyclic binary counter of \citet{disser2023unified}, every policy visited along Bland's trajectory is proper. Theorem~\ref{thm:transfer} therefore lifts its counting trajectory to a discounted two-player game on which Dantzig makes at least $2^k-1$ switches.

\subsection{Source graph and initialization}

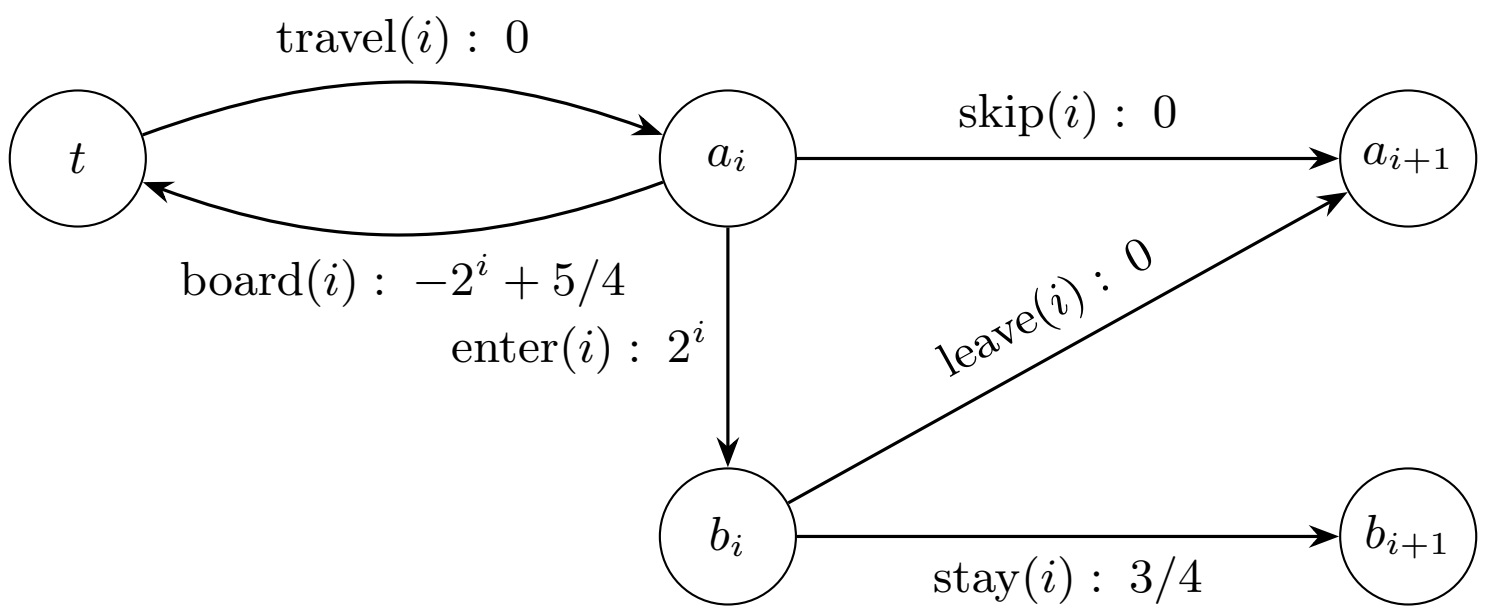
\begin{figure}[htbp]
\centering
\begin{tikzpicture}[font=\small]
\node[state] (t) at (0,0) {$t$};
\node[state] (a) at (4.3,0) {$a_i$};
\node[state] (an) at (8.8,0) {$a_{i+1}$};
\node[state] (b) at (4.3,-2.5) {$b_i$};
\node[state] (bn) at (8.8,-2.5) {$b_{i+1}$};
\draw[edge] (t) to[bend left=20] node[above] {$\trav(i):\ 0$} (a);
\draw[edge] (a) to[bend left=20] node[below] {$\brd(i):\ -2^i+5/4$} (t);
\draw[edge] (a) -- node[left] {$\ent(i):\ 2^i$} (b);
\draw[edge] (a) -- node[above] {$\skp(i):\ 0$} (an);
\draw[edge] (b) -- node[below] {$\sty(i):\ 3/4$} (bn);
\draw[edge] (b) -- node[above,sloped] {$\lev(i):\ 0$} (an);
\end{tikzpicture}
\caption{One level of $\mathcal B_k$ from \citet{disser2023unified}. Each edge shows its action and reward. At level $k$, the upper and lower successors are $s$ and $d$, respectively. Both forced actions, $d\to s$ and $s\to s$, pay zero.}\label{fig:source}
\end{figure}

In the source MDP $\mathcal B_k$, every nonterminal cycle returns to a hub state $t$ via a board action (Figure~\ref{fig:source}). For $k\ge1$, the state set consists of $t,d,s$ together with $a_i,b_i$ for $1\le i\le k$, using the conventions $a_{k+1}=s$ and $b_{k+1}=d$. The table below specifies the transitions, rewards, and Bland ranks of all $m=6k+2$ actions.

\begin{center}
\begin{tabular}{llll}
\toprule
Action & Transition & Reward & Rank\\
\midrule
$\trav(i)$ & $t\to a_i$ & $0$ & $i$\\
$\ent(i)$ & $a_i\to b_i$ & $2^i$ & $k+1+5(i-1)$\\
$\skp(i)$ & $a_i\to a_{i+1}$ & $0$ & $k+2+5(i-1)$\\
$\brd(i)$ & $a_i\to t$ & $-2^i+5/4$ & $k+3+5(i-1)$\\
$\sty(i)$ & $b_i\to b_{i+1}$ & $3/4$ & $k+4+5(i-1)$\\
$\lev(i)$ & $b_i\to a_{i+1}$ & $0$ & $k+5+5(i-1)$\\
Forced action & $d\to s$ & $0$ & $6k+1$\\
Forced action & $s\to s$ & $0$ & $6k+2$\\
\bottomrule
\end{tabular}
\end{center}

The initial policy $\sigma_0$ chooses $\trav(1)$ at $t$, $\skp(i)$ at each $a_i$, and $\lev(i)$ at each $b_i$. Every state reaches $s$ along zero-reward edges, so both undiscounted and discounted values under $\sigma_0$ vanish identically.

\begin{lemma}[Negative cycles and absorption]\label{lem:proper}
Every nonterminal simple cycle in $\mathcal B_k$ has total reward at most $-3/4$, and every policy reached from $\sigma_0$ by strictly improving undiscounted single-state switches reaches $s$ from every state within $2k+2$ transitions.
\end{lemma}
\begin{proof}
Any nonterminal simple cycle leaves $t$ via $\trav(j)$, follows level-increasing edges to $a_i$ ($j\le i$), and returns to $t$ via $\brd(i)$. Setting $U(a_h)=2^h$ and $U(b_h)=2^{h+1}$, every forward edge $e:x\to y$ satisfies $r_e\le U(y)-U(x)$, so the cycle's total reward is at most $U(a_i)-U(a_j)-2^i+5/4=5/4-2^j\le-3/4$.

Because any cycle created by a single switch from a proper policy has total reward equal to that switch's undiscounted advantage, strictly improving switches preserve properness. Induction from $\sigma_0$ shows that every visited policy reaches $s$ without repeating any of the $2k+2$ nonsink states, taking at most $2k+2$ transitions.
\end{proof}

\subsection{Canonical policies and Bland's trajectory}

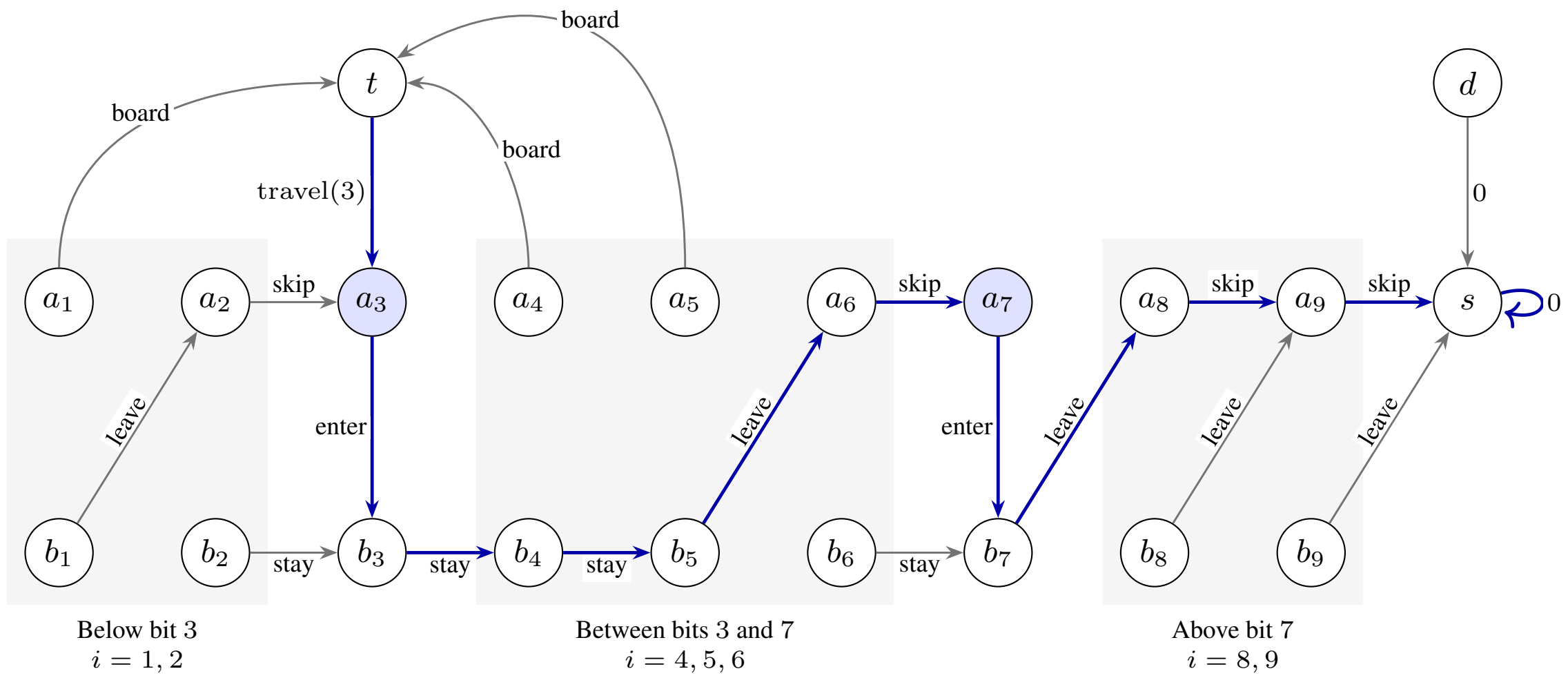
\begin{figure}[htbp]
\centering
\begin{tikzpicture}[font=\small,
  cstate/.style={circle,draw,fill=white,minimum size=6.5mm,inner sep=1pt},
  cpath/.style={-{Stealth[length=1.7mm]},draw=blue!65!black,line width=0.9pt},
  cother/.style={-{Stealth[length=1.7mm]},draw=black!55,line width=0.55pt},
  clabel/.style={font=\scriptsize,fill=white,inner sep=1pt}]
\fill[black!4] (-0.5,0.6) rectangle (2,-2.9);
\fill[black!4] (4,0.6) rectangle (8,-2.9);
\fill[black!4] (10,0.6) rectangle (12.5,-2.9);
\node[cstate] (a1) at (0.0,0) {$a_1$};
\node[cstate] (b1) at (0.0,-2.4) {$b_1$};
\node[cstate] (a2) at (1.5,0) {$a_2$};
\node[cstate] (b2) at (1.5,-2.4) {$b_2$};
\node[cstate,fill=blue!12] (a3) at (3.0,0) {$a_3$};
\node[cstate] (b3) at (3.0,-2.4) {$b_3$};
\node[cstate] (a4) at (4.5,0) {$a_4$};
\node[cstate] (b4) at (4.5,-2.4) {$b_4$};
\node[cstate] (a5) at (6.0,0) {$a_5$};
\node[cstate] (b5) at (6.0,-2.4) {$b_5$};
\node[cstate] (a6) at (7.5,0) {$a_6$};
\node[cstate] (b6) at (7.5,-2.4) {$b_6$};
\node[cstate,fill=blue!12] (a7) at (9.0,0) {$a_7$};
\node[cstate] (b7) at (9.0,-2.4) {$b_7$};
\node[cstate] (a8) at (10.5,0) {$a_8$};
\node[cstate] (b8) at (10.5,-2.4) {$b_8$};
\node[cstate] (a9) at (12.0,0) {$a_9$};
\node[cstate] (b9) at (12.0,-2.4) {$b_9$};
\node[cstate] (t) at (3,2.1) {$t$};
\node[cstate] (d) at (13.5,2.1) {$d$};
\node[cstate] (s) at (13.5,0) {$s$};
\draw[cpath] (t) -- node[clabel,left] {$\trav(3)$} (a3);
\draw[cother] (a1.north) .. controls (0,1.6) and (1.2,2.1) ..
  node[midway,clabel,above] {board} (t.west);
\draw[cother] (a4.north) .. controls (4.5,1) and (4.1,2.1) ..
  node[midway,clabel,right] {board} (t.east);
\draw[cother] (a5.north) .. controls (6,2.9) and (4.5,3.1) ..
  node[midway,clabel,above] {board} (t.north east);
\draw[cother] (a2) -- node[clabel,above] {skip} (a3);
\draw[cpath] (a3) -- node[clabel,left] {enter} (b3);
\draw[cpath] (a6) -- node[clabel,above] {skip} (a7);
\draw[cpath] (a7) -- node[clabel,left] {enter} (b7);
\draw[cpath] (a8) -- node[clabel,above] {skip} (a9);
\draw[cpath] (a9) -- node[clabel,above] {skip} (s);
\draw[cother] (b1) -- node[clabel,above,sloped] {leave} (a2);
\draw[cother] (b2) -- node[clabel,below] {stay} (b3);
\draw[cpath] (b3) -- node[clabel,below] {stay} (b4);
\draw[cpath] (b4) -- node[clabel,below] {stay} (b5);
\draw[cpath] (b5) -- node[clabel,above,sloped] {leave} (a6);
\draw[cother] (b6) -- node[clabel,below] {stay} (b7);
\draw[cpath] (b7) -- node[clabel,above,sloped] {leave} (a8);
\draw[cother] (b8) -- node[clabel,above,sloped] {leave} (a9);
\draw[cother] (b9) -- node[clabel,above,sloped] {leave} (s);
\draw[cother] (d) -- node[clabel,right] {$0$} (s);
\draw[cpath] (s) edge[loop right] node[clabel,right] {$0$} (s);
\node[font=\scriptsize,align=center] at (0.75,-3.3) {Below bit $3$\\$i=1,2$};
\node[font=\scriptsize,align=center] at (6,-3.3) {Between bits $3$ and $7$\\$i=4,5,6$};
\node[font=\scriptsize,align=center] at (11.25,-3.3) {Above bit $7$\\$i=8,9$};
\end{tikzpicture}
\caption{Canonical policy $\sigma_{68}$ in $\mathcal B_9$. The filled states mark the set bits $3$ and $7$, since $68=2^2+2^6$. All arrows show selected actions. The thick blue path starts at $t$ with $\trav(3)$, and the shaded regions mark blocks of zero bits.}\label{fig:canonical-policy}
\end{figure}

Bland visits a distinct canonical policy $\sigma_x$ for each $k$-bit integer $x=0,1,\ldots,2^k-1$ in order, with the actions $\ent(i)$ encoding the set bits of $x$ (Figure~\ref{fig:canonical-policy}).

\begin{definition}[Canonical counter policy {\citep[Definition~6]{disser2023unified}}]\label{def:canonical}
Set $\sigma_x=\sigma_0$ for $x=0$. For $0<x<2^k$, index the binary bits $1,\ldots,k$ from least significant to most significant, and modify $\sigma_0$ by selecting $\trav(q_1)$ at $t$ for the lowest set bit $q_1$ and $\ent(i)$ at $a_i$ for every set bit $i$.

In addition, for each set bit $q\ge3$, let $p$ be the largest set bit below $q$ (with $p=0$ if $q=q_1$). Whenever $q\ge p+2$, set
\[
\begin{aligned}
    \sigma_x(a_j)&=\brd(j) &&\text{for }p<j\le q-2,\\
    \sigma_x(b_j)&=\sty(j) &&\text{for }\max\{1,p\}\le j\le q-3\ \text{or }j=q-1,
\end{aligned}
\]
leaving all remaining states at their choices in $\sigma_0$.
\end{definition}

\begin{theorem}[Bland counting trajectory {\citep[Lemmas~10 and~13]{disser2023unified}}]\label{thm:source}
Starting from $\sigma_0$ under the ranks above, Bland visits the distinct policies $\sigma_0,\sigma_1,\ldots,\sigma_{2^k-1}$ in order, passing through possible intermediate policies between consecutive canonical ones and making at least $2^k-1$ switches.
\end{theorem}

\subsection{Transfer and encoding for the cyclic source}\label{sec:encoding}

Using Lemma~\ref{lem:proper} together with $|r_a|\le 2^k$ and $4r_a\in\Z$, instantiate the default bounds and parameters of Section~\ref{sec:gate-parameters} with
\begin{equation}\label{eq:instantiation}
    H=2k+2,\qquad R=2^k,\qquad D=4,\qquad p_k=64Q,\qquad \gamma_k=1-1/p_k,
\end{equation}
and denote the resulting game at discount $\gamma_k$ by $\mathcal G_k$.

\begin{proposition}[Dantzig lower bound from the cyclic counter]\label{thm:main}
Starting from $\widehat\sigma_0$, Dantzig makes at least $2^k-1$ switches on $\mathcal G_k$, with a unique choice at each step on the simulated prefix and a unique all-state stationary Min best response at every visited strategy. The same conclusions hold for any rational $\gamma$ with $(1-\gamma)^{-1}\ge p_k$ when the gate rewards are parameterized by $\gamma$. At $\gamma_k$, $\mathcal G_k$ has $8k+5$ states, $18k+6$ actions, and an $O(k^2)$-bit encoding, and lifting an optimal source policy solves $\mathcal G_k$ in strongly polynomial time.
\end{proposition}
\begin{proof}
Lemma~\ref{lem:proper} ensures that every policy along the Bland trajectory of Theorem~\ref{thm:source}, including intermediate ones, is proper, so Theorem~\ref{thm:transfer} gives the switch lower bound and uniqueness for every admissible $\gamma$. Gating all $m=6k+2$ actions alternates control and yields $N=(2k+3)+m=8k+5$ states and $M=3m=18k+6$ actions.

At $\gamma_k$, clearing denominators by a common positive factor gives
\begin{equation}\label{eq:rewardbits}
    p_k(p_k-1)\left(\frac{r_a}{\gamma_k},\,\frac{(1-\gamma_k^2)K_a}{\gamma_k}\right)
    =\bigl(p_k^2r_a,\,(2p_k-1)K_a\bigr)\in\Z^2,
\end{equation}
since $4\mid p_k$ and $4r_a,K_a\in\Z$. By~\eqref{eq:K} and~\eqref{eq:Q}, $K_{\max},p_k=O(k^2 2^k)$, so each reward uses $O(k)$ bits and the entire game uses $O(k^2)$ bits.

Finally, let $V^*$ be the optimal source value at discount $\gamma_k^2$. Evaluating $\sigma_0$ gives $V^*\ge0$. If an optimal source policy contained a nonterminal cycle $C$, summing its evaluation equations around $C$ would give cycle reward $(1-\gamma_k^2)\sum_{u\in C}V^*(u)\ge0$, contradicting Lemma~\ref{lem:proper}. Hence an optimal source policy is proper and satisfies $\|V^*\|_\infty\le HR=B_V$. For every $a:u\to w$, source optimality and $K_a>B_V$ yield
\begin{equation}\label{eq:optimal-original-branch}
    r_a+\gamma_k^2 V^*(w)\le V^*(u)<(1-\gamma_k^2)K_a+\gamma_k^2 V^*(u),
\end{equation}
so Min takes the forward branch at every gate, and $W^*(u)=V^*(u)$ together with $W^*(g_a)=r_a/\gamma_k+\gamma_k V^*(w)$ satisfies $\mathcal T W^*=W^*$. Solving the deterministic source MDP in strongly polynomial time~\citep{post2015simplex} and lifting the resulting optimal policy solves $\mathcal G_k$.
\end{proof}

\section{Proofs of the upper bounds and solver extensions}\label{app:component-proofs}

Once all downstream components have been solved, every exit from the current component has a known optimal payoff. The component solver uses those boundary payoffs to eliminate suboptimal Min exits permanently, whereas the global switch upper bounds track how often downstream switches can alter boundary payoffs. The final two subsections establish the occupation-based switch bound and the branching extension beyond $\mathcal C_1$.

\subsection{Exit elimination with fixed boundaries}\label{app:exit-elimination}

Forcing Min to exit wherever possible produces a deterministic Max MDP whose value upper-bounds the local game value, so any exit beaten by continuing even under that upper bound can be discarded permanently.

\begin{lemma}[Exit elimination with fixed boundaries]\label{lem:fixed-exit-elimination}
Consider a local game with $n$ states outside a forced zero-reward sink. Assume every Min state has at most one non-exit action and preprocessing has retained a single least-payoff exit at each Min state that has exits. Among the $n^{\Min}$ nonsink Min states, let $e$ possess both an exit and a non-exit action, and let $m^{\Max}$ be the number of Max actions outside the sink. Then at most $e+1$ Max MDP solves find the local optimal values and strategies, using
\begin{align}\label{eq:exit-elimination-cost}
    O\!\left((e+1)(n+1)(m^{\Max}+n^{\Min}+1)\right)
\end{align}
arithmetic operations.
\end{lemma}
\begin{proof}
Initialize $\tau$ as in Section~\ref{sec:component-solver}, let $U^\tau$ be the optimal value of the resulting Max MDP, and write $w^*$ for the local game value, so $w^*\le U^\tau$. At each Min state $x$ where $\tau$ currently takes its retained exit of payoff $v_x$ and also has a non-exit action $a_x$, delete the exit whenever
\begin{align}\label{eq:permanent-exit-test}
    c(a_x)+\gamma U^\tau(f(a_x))<v_x.
\end{align}
Because $w^*\le U^\tau$, replacing $U^\tau$ by $w^*$ preserves this strict inequality, proving that every deleted exit is strictly suboptimal at $w^*$. Switching $\tau$ to $a_x$ at all such states maps $U^\tau$ downward under the new MDP Bellman operator and yields a new upper bound $U^{\tau'}\le U^\tau$.

When no remaining exit satisfies~\eqref{eq:permanent-exit-test}, $U^\tau(x)=v_x$ attains the Bellman minimum at every two-action Min state while MDP optimality satisfies all other Bellman equations, giving $U^\tau=w^*$ and an optimal local strategy pair. Every iteration before termination deletes at least one of the $e$ eligible exits. Including the sink, each Max MDP has $n+1$ states and $m^{\Max}+n^{\Min}+1$ actions, so solving each MDP in $O(nm)$ arithmetic operations via \citet{madani2010discounted} together with exit tests and strategy recovery gives~\eqref{eq:exit-elimination-cost}.
\end{proof}

\subsection{Proof of Theorem~\ref{thm:component-solver}}\label{app:component-solver-proof}

\begin{proof}
Processing SCCs in reverse topological order ensures that $W^*(y)$ is already known at every exit $a:x\to y$ from the current component $C$, so the boundary substitution~\eqref{eq:fixed-boundary} preserves the Bellman equations on $C$. Let $n_C$ and $m_C$ denote the numbers of states and actions in $C$. A single-player component requires one deterministic MDP solve in $O((n_C+1)(m_C+1))$ operations, whereas a mixed component has $e\le n_C-1$ and by Lemma~\ref{lem:fixed-exit-elimination} requires at most $n_C$ solves of $O((n_C+1)(m_C+1))$ operations each. Summing over all components via $\sum_C n_C=N$ and $\sum_C m_C=M$, together with $O(N+M)$ work for SCC decomposition, exit preprocessing, and tight-action recovery, gives $O(N^2M)$ arithmetic operations in total.

Fixing optimal strategies in all solved downstream components identifies each intermediate MDP value vector on $C$ with the value of a stationary strategy pair in the original game. Because its discounted path-and-cycle evaluation uses at most $N$ distinct states, every local value and boundary payoff has bit length polynomial in the input size, as do all intermediate numbers inside the solver of \citet{madani2010discounted}.
\end{proof}

\textbf{Two independent ko fights.} Two separated single-stone ko sites in Go illustrate how a mixed SCC can naturally satisfy the $\mathcal C_1$ condition. At each site, either player may capture the opposing ko stone or pass, subject to a ban on immediate recapture while allowing later repetitions, with two consecutive passes ending play. With surrounding stones fixed, a game state records the color of the ko stone at each of the two sites, the player to move, and the previous move. Assign bounded rational action rewards and a common discount $\gamma$. The diagram below depicts one of the two sites.

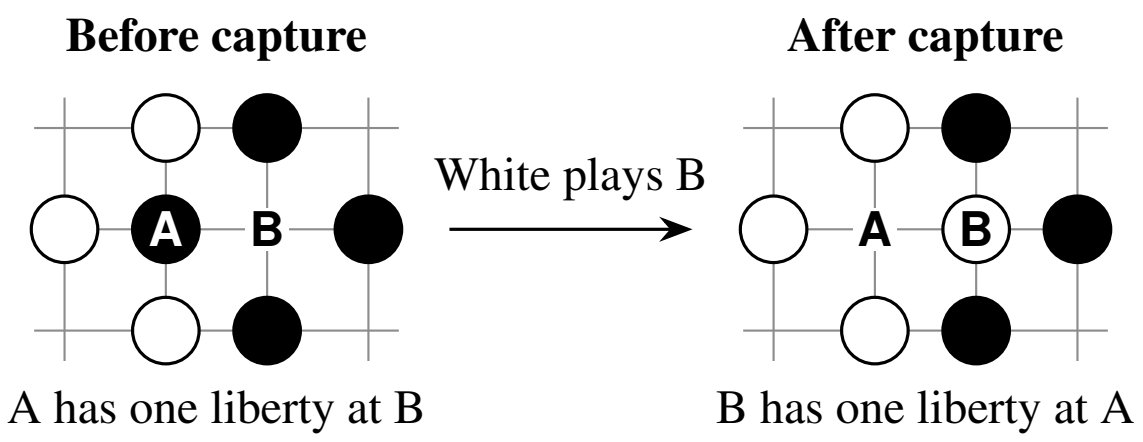
\begin{center}
\begin{tikzpicture}[x=0.67cm,y=0.67cm,font=\small,
    stone/.style={circle,draw=black,line width=0.6pt,minimum size=0.44cm,inner sep=0pt},
    marked/.style={font=\small\bfseries\sffamily,inner sep=0.8pt}]
  \foreach \offset in {0,7} {
    \begin{scope}[shift={(\offset,0)}]
      \foreach \i in {0,...,3} \draw[black!45] (\i,-0.30) -- (\i,2.30);
      \foreach \j in {0,...,2} \draw[black!45] (-0.30,\j) -- (3.30,\j);
      \foreach \p in {(2,0),(2,2),(3,1)} \node[stone,fill=black] at \p {};
      \foreach \p in {(1,0),(1,2),(0,1)} \node[stone,fill=white] at \p {};
    \end{scope}
  }
  \node[stone,fill=black,text=white,marked] at (1,1) {A};
  \node[marked,fill=white,text=black] at (2,1) {B};
  \node[marked,fill=white,text=black] at (8,1) {A};
  \node[stone,fill=white,text=black,marked] at (9,1) {B};
  \node[text=black,font=\small\bfseries] at (1.5,2.88) {Before capture};
  \node[text=black,font=\small\bfseries] at (8.5,2.88) {After capture};
  \draw[-{Stealth[length=2.2mm]},black,semithick] (3.80,1) -- (6.20,1);
  \node[text=black,align=center] at (5,1.50) {White plays B};
  \node[text=black] at (1.5,-0.80) {A has one liberty at B};
  \node[text=black] at (8.5,-0.80) {B has one liberty at A};
\end{tikzpicture}
\end{center}

Start with Black holding the ko stone at A at both sites and White to move, just after a Black capture on the left. White captures on the right, Black is forced by the recapture ban to pass, and White captures on the left. Repeating these three moves with colors swapped restores the exact initial state. Passing instead of either left capture ends play immediately (as the second consecutive pass), whereas passing instead of either right capture leaves the opponent only a terminal pass. Each of the six cycle states therefore has a unique internal action, placing the reachable game in $\mathcal C_1$. (In tournament Go, Japanese rules supplement this local repetition model with dedicated double-ko seki confirmation rules~\citep{japanesego1989}.)

\subsection{Response changes}\label{app:response-phases}

Whenever a single Min strategy remains an all-state best response prior to every switch in a contiguous segment of a Dantzig trajectory, those switches share the exact advantages of one deterministic MDP.

\begin{lemma}[Dantzig switches in fixed-response phases]\label{lem:response-phases}
If the switches of a Dantzig trajectory partition into $q$ consecutive phases such that each phase admits one stationary Min strategy that is an all-state best response before every switch in that phase, then
\begin{align}\label{eq:response-phases}
    T\le qP(N,M).
\end{align}
\end{lemma}
\begin{proof}
Fixing that phase's common best response produces a deterministic Max MDP in which every pre-switch Max strategy has the same value vector, and hence the same action advantages, as in the two-player game. The switches in the phase (including its final switch, even if that switch alters Min's subsequent best response) therefore form a valid Dantzig trajectory prefix in that MDP and number at most $P(N,M)$.
\end{proof}

\subsection{Proof of Theorem~\ref{thm:fixed-exits}}\label{app:fixed-exits-proof}

\begin{proof}
At each Min state $x$ that has sink actions, fix one of least immediate reward $v_x$. Whenever $x$ also has a nonsink action $a_x$, select that sink action if and only if
\begin{align}\label{eq:exit-test}
    v_x\le c(a_x)+\gamma W^\sigma(f(a_x)),
\end{align}
and select $a_x$ otherwise. At Min states with only sink actions, always select the fixed least-reward sink action, while all remaining Min states have a single forced action. These choices attain every Bellman minimum and hence define an all-state best response at each step. Because $W^\sigma$ is componentwise nondecreasing along the trajectory (Lemma~\ref{lem:strict}), once~\eqref{eq:exit-test} holds at $x$ it continues to hold at all later steps, including on ties. Consequently, only the $e$ states with both choices can ever change action, each at most once, partitioning the trajectory into at most $e+1$ fixed-response phases and giving $T\le(e+1)P(N,M)$ by Lemma~\ref{lem:response-phases}. Computing exact responses via the strongly polynomial deterministic-MDP solver of \citet{madani2010discounted} yields strong polynomiality.
\end{proof}

\paragraph{Response reversals when an exit payoff changes.}

Let $\gamma=1/2$ and let $s$ be a forced zero-reward sink. Max chooses between $u\to x$ (reward $2$) and $u\to s$ (reward $5$), and at $z$ chooses between two sink actions $z\to s$ with rewards $0$ and $4$. At $x$, Min chooses between continuing to $u$ and exiting to $z$, both with reward $0$. Starting from $u\to x$ and the zero-reward action at $z$, the initial advantages at $z$ and $u$ are $4$ and $3$, so global Dantzig switches first at $z$ and then at $u$.

\begin{center}
\begin{tabular}{lcccc}
\toprule
Strategy & $W(u)$ & $W(z)$ & $W(x)$ & Response at $x$\\
\midrule
Initial & $2$ & $0$ & $0$ & exit to $z$\\
After changing $z$ & $8/3$ & $4$ & $4/3$ & continue to $u$\\
After changing $u$ & $5$ & $4$ & $2$ & exit to $z$\\
\bottomrule
\end{tabular}
\end{center}

Within the mixed SCC $\{u,x\}$, the downstream switch at $z$ raises the exit payoff $\gamma W(z)$ from $0$ to $2$, causing Min at $x$ to switch from exiting to continuing, before the local switch at $u$ raises $\gamma W(u)$ to $5/2$ and makes exiting optimal again.

\subsection{Proofs of the condensation-height bounds}\label{app:height-proof}

Because only switches strictly downstream of a component $C$ can alter its exit payoffs, bounding local switches per fixed-boundary phase yields a recurrence over the condensation graph. All bounds below hold for every finite trajectory prefix.

\begin{lemma}[Local switches between downstream updates]\label{lem:local-phase-budget}
Let $T_C$ count Max switches inside $C$ and let $\operatorname{Desc}(C)$ denote the set of components strictly downstream of $C$. If $C$ makes at most $L_C$ switches per fixed-boundary phase, then
\begin{align}\label{eq:component-recurrence}
    T_C\le L_C\left(1+\sum_{B\in\operatorname{Desc}(C)}T_B\right).
\end{align}
For any rule that executes strictly improving single-state Max switches under exact responses,~\eqref{eq:component-recurrence} holds with $L_C=S_C-1$, where $S_C$ is defined in~\eqref{eq:local-strategy-count}.
\end{lemma}
\begin{proof}
Splitting the trajectory at every switch in $\operatorname{Desc}(C)$ produces at most $1+\sum_{B\in\operatorname{Desc}(C)}T_B$ fixed-boundary phases. Within each phase, the downstream Max strategy is constant, fixing every exit payoff $c(a)+\gamma W^\sigma(y)$. Redirecting every exit from $C$ separately to a local zero-reward sink preserves all local Max strategies, values, and advantages on $C$, regardless of switches outside $C\cup\operatorname{Desc}(C)$.

With fixed boundary payoffs, each local Max strategy on $C$ determines a unique local value vector, and each local positive-advantage switch strictly increases the switched state's value while decreasing none (Lemma~\ref{lem:strict}). No local strategy can repeat within a phase, limiting each phase to at most $S_C-1$ local switches.
\end{proof}

\begin{proof}[Proof of Theorem~\ref{thm:component-height}]
Every global Dantzig switch occurring in $C$ also maximizes local advantage on $C$. For a mixed component $C$ of a game in $\mathcal C_1$, let $e_C$ count Min states in $C$ having both an internal action and an exit. Applying Theorem~\ref{thm:fixed-exits} to the fixed-boundary local game together with Lemma~\ref{lem:local-phase-budget} allows
\begin{align}\label{eq:height-local-budget}
    L_C=\min\{S_C-1,(e_C+1)P(n_C+1,m_C+1)\}.
\end{align}
The same formula applies to Max-only components with $e_C=0$, while Min-only components satisfy $L_C=0$. Using $e_C\le n_C$ and $P(n,m)=O(n^2m^2\log^2(n+1))$, summing over all components gives
\begin{equation}\label{eq:component-height-budget}
    L=\sum_C L_C=O(N^3M^2\log^2(N+1)).
\end{equation}

Let $\operatorname{ht}(C)$ be the maximum number of vertices on a directed path in the condensation graph starting at $C$, let $L_j$ sum $L_C$ over components with $\operatorname{ht}(C)=j$, and let $U_j$ count switches in all components with $\operatorname{ht}(C)\le j$ (setting $U_0=0$). Because every $B\in\operatorname{Desc}(C)$ has $\operatorname{ht}(B)<\operatorname{ht}(C)$,~\eqref{eq:component-recurrence} yields $U_j-U_{j-1}\le L_j(1+U_{j-1})$. Iterating from $j=1$ to $\rho$ and applying the arithmetic--geometric mean inequality gives
\[
    1+T=1+U_\rho\le\prod_{j=1}^{\rho}(1+L_j)
       \le(1+L/\rho)^\rho.
\]
Because this holds for every finite prefix, it bounds total switches to termination. For fixed $\rho$, the bound is polynomial in $N$ and $M$, and pairing global Dantzig with a strongly polynomial deterministic-MDP response solver makes the overall algorithm strongly polynomial.
\end{proof}

Unrolling~\eqref{eq:component-recurrence} directly along reachability chains without grouping by height yields the sharper bound
\begin{align}\label{eq:reachable-chain-bound}
    T\le
    \sum_{\substack{k\ge0,\ C_0\succ C_1\succ\cdots\succ C_k}}
       \prod_{i=0}^{k}L_{C_i},
\end{align}
where $C\succ B$ means $B\in\operatorname{Desc}(C)$, counting each nonempty chain once even when multiple condensation paths connect its elements (and bounded above by $\prod_C S_C-1$ when $L_C=S_C-1$).

\begin{proof}[Proof of Theorem~\ref{thm:bounded-components}]
For each component $C$, let $R_C=T_C+\sum_{B\in\operatorname{Desc}(C)}T_B$ count switches in $C$ and its strict descendants, and let $\operatorname{Succ}(C)$ denote its direct successor components in the condensation graph. Because $\operatorname{Desc}(C)$ is covered by $\bigcup_{B\in\operatorname{Succ}(C)}(\{B\}\cup\operatorname{Desc}(B))$, substituting $L_C=S_C-1$ into~\eqref{eq:component-recurrence} gives
\begin{align}\label{eq:height-region-recurrence}
    R_C\le(S_C-1)+S_C\sum_{B\in\operatorname{Succ}(C)}R_B.
\end{align}
Sink components satisfy $R_C\le S_C-1$, and every other component has at most $d_{\mathrm{out}}$ direct successors, each of strictly smaller height, so induction on $\operatorname{ht}(C)$ yields
\[
    R_C\le(S_{\max}-1)\sum_{j=0}^{\operatorname{ht}(C)-1}(d_{\mathrm{out}}S_{\max})^j.
\]
Because every component is reachable from at least one of the $r$ source components, summing $R_C$ over those $r$ sources bounds $T$ and proves~\eqref{eq:bounded-components}. When $n_C\le n_{\max}$ and $|\A(x)|\le d_{\max}$, the estimates $S_{\max}\le d_{\max}^{n_{\max}}$, $d_{\mathrm{out}}\le n_{\max}d_{\max}$, and $r\le N$ yield $T\le N2^{O(\rho)}$ for fixed $n_{\max},d_{\max}$. When $\rho\le c_\rho\log(N+1)$ for fixed $c_\rho$, $T$ is polynomial in $N$, and each Dantzig iteration performs at most $M$ advantage comparisons plus one strongly polynomial deterministic-MDP response solve.
\end{proof}

\subsection{Proof of Theorem~\ref{thm:response-occupation} and Corollary~\ref{cor:transient-decisions}}\label{app:response-occupation-proof}

Whereas the MDP bounds of \citet{scherrer2016improved} require uniform occupation bounds across policies, the two-player bound below requires a visitation bound for only one optimal Min response at each visited Max strategy.

\begin{proof}
If $J=\varnothing$, then $T=0$. Otherwise write $W_t=W^{\sigma_t}$, $e_t=W^*-W_t\ge0$, and $E_t=\sum_{z\in X} e_t(z)$. At any nonoptimal step $t$, let $g_t>0$ be the largest positive Max advantage. Because $\mathcal T W_t-W_t$ vanishes on $X_{\Min}$ and has sup norm $g_t$, $\gamma$-contraction of $\mathcal T$ gives $\|e_t\|_\infty\le\gamma\|e_t\|_\infty+g_t$, and hence
\begin{align}\label{eq:occupation-error-residual}
    \|e_t\|_\infty\le hg_t,\qquad E_t\le Nhg_t.
\end{align}
By Lemma~\ref{lem:strict}, a Dantzig switch increases the switched state's value by at least $g_t$ and lowers none, while modified simplex maximizes the exact total gain and therefore also increases $F$ by at least $g_t$ (with positive advantage by Lemma~\ref{lem:gain-sign}). Both rules thus satisfy
\begin{align}\label{eq:occupation-contraction}
    E_{t+1}\le E_t-g_t
    \le\left(1-\frac1{Nh}\right)E_t.
\end{align}

Define the optimality loss of each action $a$ by
\begin{align}\label{eq:occupation-optimal-loss}
    \Delta^*(a)=W^*(\src(a))-c(a)-\gamma W^*(f(a)),
\end{align}
which is nonnegative on $X_{\Max}$ (zero at forced Max states and on at least one optimal action per state) and nonpositive on $X_{\Min}$. For any stationary pair $\pi$, let $a_j^z$ denote the action taken at step $j$ from initial state $z$, and define the discounted occupation measure
\begin{equation}\label{eq:response-occupation}
    x^\pi(a)=\sum_{z\in X}\sum_{j=0}^{\infty}\gamma^j\mathbf{1}_{\{a_j^z=a\}}.
\end{equation}
At iteration $t$, choose an all-state best response $\tau_t$ satisfying the assumed visitation bound and set $\pi_t=(\sigma_t,\tau_t)$, so that
\begin{equation}\label{eq:response-occupation-budget}
    \sum_{u\in J}x^{\pi_t}(\sigma_t(u))\le B.
\end{equation}
Along the play $(z_0,a_0^z,z_1,\ldots)$ of $\pi_t$ from $z_0=z$, policy evaluation gives $e_t(z_j)=\Delta^*(a_j^z)+\gamma e_t(z_{j+1})$. Iterating and letting the bounded tail vanish, then summing over $z\in X$ and dropping nonpositive terms, yields
\begin{align}\label{eq:occupation-loss-budget}
    E_t=\sum_a x^{\pi_t}(a)\Delta^*(a)
    \le\sum_{u\in J}x^{\pi_t}(\sigma_t(u))\Delta^*(\sigma_t(u))
    \le B\max_{u\in J}\Delta^*(\sigma_t(u)).
\end{align}

Whenever $E_t>0$, some currently selected action $a=\sigma_t(u)$ ($u\in J$, $v=f(a)$) therefore satisfies $\Delta^*(a)\ge E_t/B>0$. At any step $s\ge t$ with $\sigma_s(u)=a$, evaluating $\pi_s$ at $u$ gives
\begin{align}\label{eq:occupation-persistent-loss}
    e_s(u)=\Delta^*(a)+\gamma e_s(v)\ge\Delta^*(a),\qquad E_s\ge\Delta^*(a).
\end{align}
Unless the algorithm stops earlier, after $\ell=\lceil Nh\log(2B)\rceil$ switches~\eqref{eq:occupation-contraction} gives $E_{t+\ell}\le e^{-\ell/(Nh)}E_t\le E_t/(2B)<\Delta^*(a)$. Monotonicity of $E_s$ together with~\eqref{eq:occupation-persistent-loss} permanently excludes $a$ from step $t+\ell$ onward. Because at most $\sum_{u\in J}(|\A(u)|-1)$ positive-loss Max actions exist and~\eqref{eq:occupation-loss-budget} forces $E_s=0$ once all of them are excluded,~\eqref{eq:response-occupation-bound} follows.

Under Corollary~\ref{cor:transient-decisions}, revisiting a state of $J$ along a deterministic play would create a directed cycle meeting $J$, so each of the $N$ plays visits each state of $J$ at most once and $B=N|J|\le N^2$ yields~\eqref{eq:transient-decisions-bound}. Likewise, if every play reaches an absorbing component of forced zero-reward actions within $H$ steps, one may take $B=N\sum_{j=0}^{H-1}\gamma^j\le N\min\{H,h\}$.
\end{proof}

\subsection{Arithmetic complexity of internal-action enumeration}\label{sec:branching-solver}

When a Min state in a mixed SCC has multiple internal actions, fixing one internal choice per Min state reduces that component to $\mathcal C_1$ and allows permanent exit elimination. Enumerating these internal choices independently within each mixed SCC combines the backward component framework of \citet{mei2026simple} with the deterministic-MDP solver of \citet{madani2010discounted}.

For any mixed SCC $C$ of the full action graph, let $B_C$ be the set of Min states in $C$ with at least two internal actions, and define
\begin{align}\label{eq:internal-branching}
    \deg_C(x)&=|\{a\in\A(x):f(a)\in C\}|,\qquad x\in C,\notag\\
    B_C&=\{x\in C\cap X_{\Min}:\deg_C(x)\ge2\},\qquad
    b_C=|B_C|,\qquad
    \Pi_C=\prod_{x\in B_C}\deg_C(x),
\end{align}
with empty products equal to one. Write $n_C^{\Min}=|C\cap X_{\Min}|$ and let $e_C$ count Min states in $C$ having an exit. Strong connectivity ensures $\deg_C(x)\ge1$ for all $x\in C\cap X_{\Min}$. After retaining one best exit per state that has exits, the number of retained Max actions in $C$ is
\begin{align}\label{eq:max-component-actions}
    m_C^{\Max}=\sum_{x\in C\cap X_{\Max}}
       \left(\deg_C(x)+\mathbf1_{\{x\text{ has an exit from }C\}}\right).
\end{align}

\begin{theorem}[Arithmetic bound for component enumeration]\label{thm:branching-arithmetic}
For any game, internal-action enumeration computes $W^*$ and an optimal stationary strategy pair in $O(\Pi_{\max}N^2M)$ arithmetic operations, where $\mathcal S_{\mathrm{mix}}$ denotes the set of mixed SCCs and
\begin{align}\label{eq:max-branch-product}
    \Pi_{\max}=\max\bigl(\{1\}\cup\{\Pi_C:C\in\mathcal S_{\mathrm{mix}}\}\bigr).
\end{align}
This operation bound is independent of $\gamma$ and the reward magnitudes, and all intermediate rational numbers have polynomial bit length in the input size.
\end{theorem}
\begin{proof}
Process the SCCs in reverse topological order as in Section~\ref{sec:component-solver}, retaining a single best exit wherever exits exist. Inside a mixed component $C$, each of the $\Pi_C$ branches $\eta$ starts from that same set of retained exits and keeps one internal action per Min state. Writing $W^\eta$ for the branch's local game value and $W^C$ for the unrestricted local value, restricting Min gives $W^C\le W^\eta$. Conversely, given an all-state optimal local Min strategy, the branch $\eta^*$ that keeps its chosen internal action wherever it continues (and any internal action wherever it exits) attains
\begin{align}\label{eq:branch-minimum}
    W^{\eta^*}=W^C=\min_\eta W^\eta
\end{align}
simultaneously at all states of $C$.

Solving each branch $\eta$ via Lemma~\ref{lem:fixed-exit-elimination} (restoring deleted exits before the next branch) and maintaining the coordinatewise minimum takes $O(n_C+m_C^{\Max})$ overhead per branch, absorbed into the solve bound. Writing $\mathcal S_{\mathrm{pure}}$ for the single-player SCCs and including $O(N+M)$ preprocessing and tight-action recovery gives total arithmetic cost
\begin{align}\label{eq:branching-arithmetic}
    \begin{aligned}
    T_{\mathrm{arith}}=O\!\Bigl(&N+M+\textstyle\sum_{C\in\mathcal S_{\mathrm{pure}}}(n_C+1)(m_C+1)\\
    &+\textstyle\sum_{C\in\mathcal S_{\mathrm{mix}}}\Pi_C(e_C+1)(n_C+1)(m_C^{\Max}+n_C^{\Min}+1)\Bigr).
    \end{aligned}
\end{align}
Because $e_C+1\le n_C$ and $m_C^{\Max}+n_C^{\Min}\le m_C$, each mixed component costs $O(\Pi_C n_C^2m_C)$, summing to $O(\Pi_{\max} N^2M)$. As in the proof of Theorem~\ref{thm:component-solver}, fixing optimal downstream strategies identifies every branch value with a stationary-pair value of the original game, ensuring polynomial intermediate bit lengths throughout.
\end{proof}

\begin{corollary}[Bounded internal branching]\label{cor:fixed-internal-branching}
If $b_C\le\kappa$ in every mixed SCC, the arithmetic cost is $O(N^2M)$ when $\kappa=0$ and
\begin{align}\label{eq:general-branching-bound}
    O\!\left(N^2M\left(1+\frac{M-N}{\kappa}\right)^\kappa\right)
\end{align}
when $\kappa\ge1$, giving a strongly polynomial solver for every fixed $\kappa$.
\end{corollary}
\begin{proof}
When $\kappa=0$, $\Pi_{\max}=1$. For $\kappa\ge1$, pad the product $\Pi_C$ with ones to $\kappa$ factors, whose sum satisfies $\kappa+\sum_{x\in B_C}(\deg_C(x)-1)\le\kappa+m_C-n_C\le\kappa+M-N$ because $|\A(x)|\ge1$ everywhere. The arithmetic--geometric mean inequality gives $\Pi_C\le(1+(M-N)/\kappa)^\kappa$, and Theorem~\ref{thm:branching-arithmetic} yields~\eqref{eq:general-branching-bound}.
\end{proof}

\begin{corollary}[Two-action games]\label{cor:binary-internal-branching}
If every state has at most two actions and $b=\max(\{0\}\cup\{b_C:C\in\mathcal S_{\mathrm{mix}}\})$, the total arithmetic cost is $O(2^bN^3)$, with each mixed component $C$ taking
\begin{align}\label{eq:binary-component-cost}
    O\!\left(2^{b_C}(e_C+1)(n_C+1)(m_C-b_C-e_C+1)\right)
\end{align}
operations. In particular, for any fixed $\eta\ge0$, the condition $b\le\eta\log_2 N$ gives a strongly polynomial bound of $O(N^{\eta+3})$.
\end{corollary}
\begin{proof}
Every state in $B_C$ has $\deg_C(x)=2$, so $\Pi_C=2^{b_C}$. Beyond one internal action per Min state in $C$, Min has $b_C$ extra internal actions and $e_C$ exits, whereas every Max state in $C$ has an internal action and hence at most one exit, so all Max actions in $C$ are retained. Thus $m_C^{\Max}+n_C^{\Min}=m_C-b_C-e_C$, proving~\eqref{eq:binary-component-cost}, and $M\le2N$ together with $2^b\le N^\eta$ yields the remaining bounds.
\end{proof}

\FloatBarrier
\bibliographystyle{plainnat}
\bibliography{references}
\end{document}